\documentclass[12pt]{iopart}

\newcommand{\ket}[1]{\ensuremath{|#1\rangle}}

\newcommand{\dm}[2]{\ensuremath{|#1\rangle\langle #2|}}

\newcommand{\E}{\ensuremath{\mathcal{E}}\ }
\newcommand{\Er}{\ensuremath{\mathcal{E}(\rho)}}
\newcommand{\re}{\ensuremath{\rho_{\mathcal{E}}}}
\newcommand{\x}{\ensuremath{\chi} }
\newcommand{\xt}{\ensuremath{\tilde{\chi}} }
\newcommand{\xx}{\ensuremath{\chi}}
\newcommand{\xxt}{\ensuremath{\tilde{\chi}}}

\newcommand{\ur}{\ensuremath{\underline{\ket{\rho}}}}

\newcommand{\hh}[1]{\ensuremath{\hat{\hat{#1}}}}
\newcommand{\hhh}{\ensuremath{\hat{\hat{\mathcal{H}}}}}
\newcommand{\hhp}{\ensuremath{\hat{\hat{\mathcal{P}}}}}
\newcommand{\hhr}{\ensuremath{\hat{\hat{\mathcal{R}}}}}
\newcommand{\hhre}{\ensuremath{\hat{\hat{\mathcal{R}}}_{RE}}}

\newcommand{\gks}{\ensuremath{a_{\alpha \beta}}}
\newcommand{\tgks}{\ensuremath{\widetilde{a_{\alpha \beta}}}}
\newcommand{\urt}[1]{\ensuremath{\underline{\ket{\rho(#1)}}}}
\usepackage{iopams}
\usepackage{epsfig}
\usepackage[dvips]{color}
\usepackage{graphicx}
\usepackage{subfigure}
\usepackage{amssymb}
\usepackage{amsfonts}
\begin{document}

\title[Quantum process tomography of a single solid state qubit]{Quantum process tomography of a single solid state qubit}

\author{M Howard$^1$, J Twamley$^2$, C Wittmann$^3$, T Gaebel$^3$, F Jelezko$^3$, and J Wrachtrup$^3$}

\address{$^1$Dept. of Physics, University of California, Santa Barbara, CA 93106, U.S.A.\\ $^2$Centre for Quantum Computer Technology, Macquarie
University, Sydney, New South Wales 2109, Australia\\$^3$University of Stuttgart, 3. Physical Institute, Stuttgart, Germany.}
\ead{jtwamley@ics.mq.edu}

\begin{abstract}
We present an example of quantum process tomography (QPT) performed on a single solid state qubit. The qubit used is two energy levels of the
triplet state in the Nitrogen-Vacancy defect in Diamond. Quantum process tomography is applied to a qubit which has been allowed to decohere for
three different time periods. In each case the process is found in terms of the $\chi$ matrix representation and the affine map representation.
The discrepancy between experimentally estimated process and the closest physically valid process is noted. The results of QPT performed after
three different decoherence times are used to find the error generators, or Lindblad operators, for the system, using the technique introduced
by Boulant \textit{et al.} \cite{havel}.
\end{abstract}


\section{Introduction}
Quantum process tomography (QPT) is a method of experimentally determining the unknown dynamics of a quantum system. This is crucial as a method
of checking the functionality of would-be quantum information processing devices. Using knowledge obtained from QPT applied to such a device,
one can ideally locate and possibly rectify any sources of errors or decoherence. The standard QPT technique, which involves the use of multiple
test states, was first developed by Nielsen and Chuang \cite{jmodopt} and Poyatos, Cirac and Zoller \cite{Poyatos}. A different technique,
exploiting entangled states, was subsequently proposed by Leung \cite{leungthesis} and D'Ariano and Lo Presti \cite{D'Ariano}. More recently,
methods have been developed to determine a process from a tomographically incomplete set of measurements \cite{incomplete}, as well as
techniques to ascertain the master equation describing time evolution of the system \cite{havel,superoperators}. QPT has been performed in
liquid NMR implementations \cite{nielsenNMR,Childs,Weinstein}, numerous optical systems \cite{Lo  Presti,nambu,De
Martini,cnotqpt,Mitchell,Altepeter}, atoms in an optical lattice \cite{steinberg} and in bulk solid state NMR \cite{kampermann}. To the best of
our knowledge, this work contains the first example of QPT performed on a single solid state qubit.


\section{The Nitrogen Vacancy Defect in Diamond}
The nitrogen vacancy defect (or N-V center) is a naturally occurring defect in diamond with nitrogen impurities. It can be manufactured in type
IB synthetic diamond by irradiation and subsequent annealing at temperatures above 550 C; radiation damage causes vacancies in the diamond
lattice and annealing leads to migration of vacancies towards the nitrogen atoms (see figure 1(a)). The N-V center can also be produced in type
IIA diamonds by N$^+$ ion implantation \cite{newref1,newref2}.

\begin{figure}[!ht]
    \begin{center}
    \subfigure{\epsfig{file=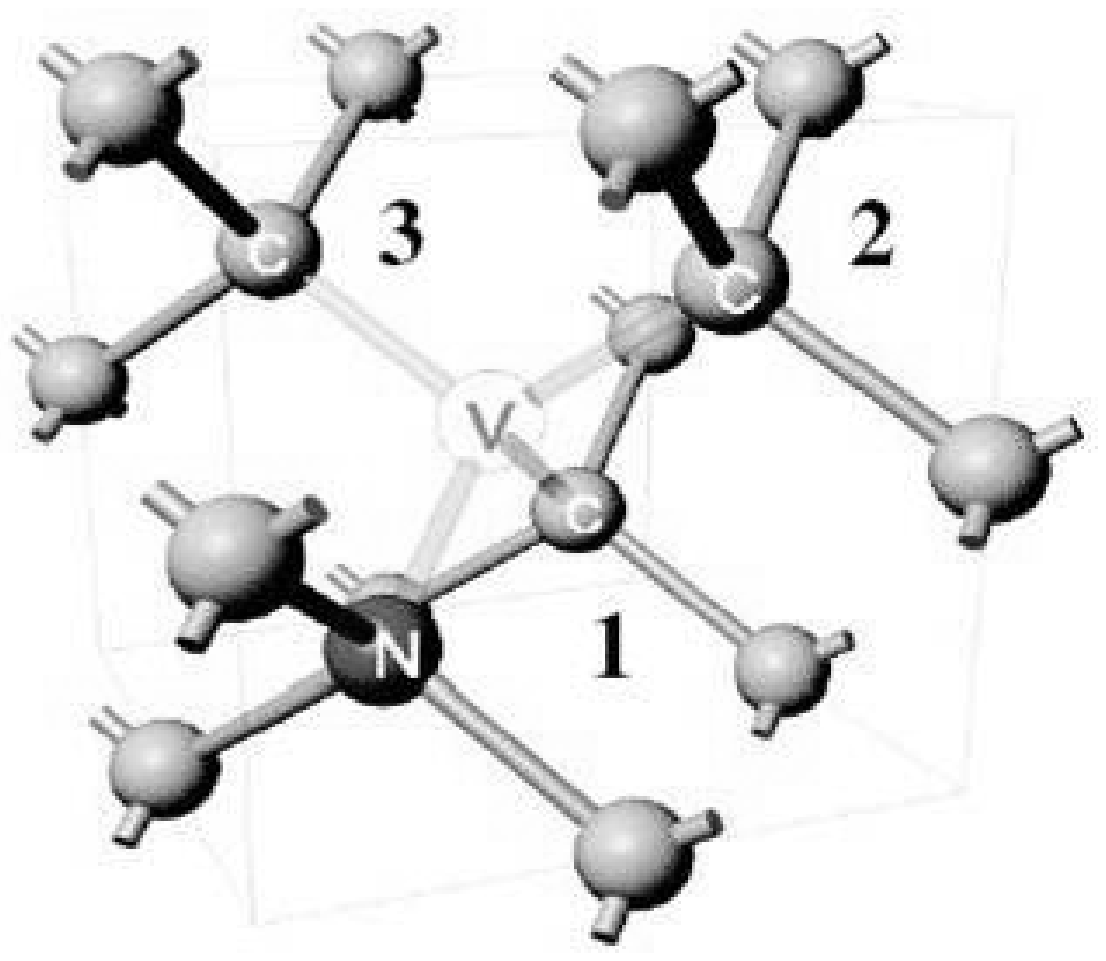, bb= 2 2 353 301, clip= ,height=43mm}}{{(a)}}
    \vspace{10mm} \subfigure{\epsfig{file=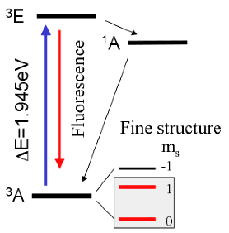, height=43mm}}{{(b)}}
  \caption{\label{levels}(a) Structure of the N-V defect (N=Nitrogen, V=Vacancy and C=Carbon) (b) Energy level scheme of N-V defect
   in a weak magnetic field.  $^3E$ and $^3A$ are excited and ground triplet states respectively. The $m_s=0$ and $m_s=1$ states are used as a logical qubit }
\end{center}
\end{figure}
\begin{figure}[!ht]
    \begin{center}
    \subfigure{\epsfig{file=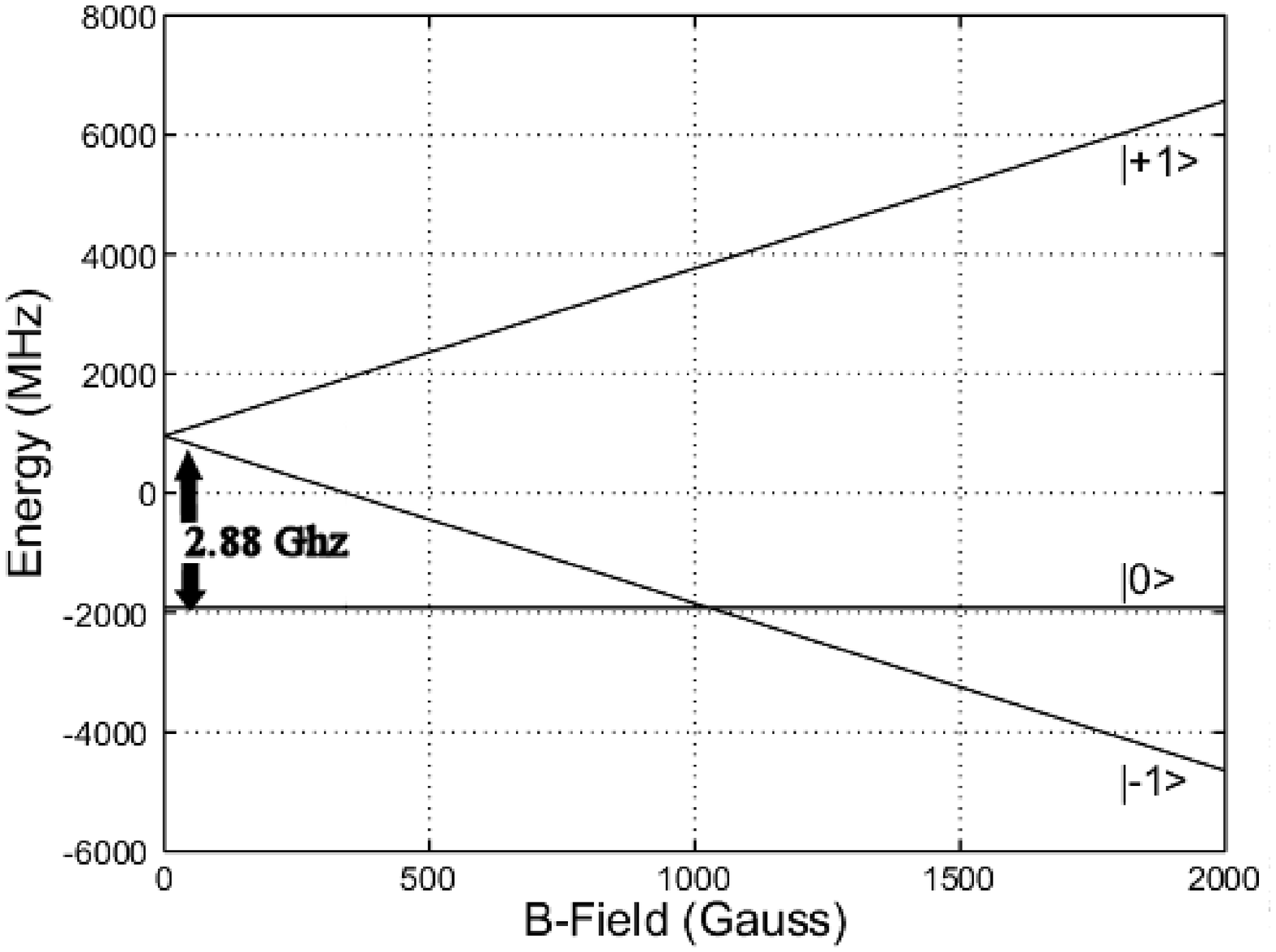, height=50mm}}{{(a)}}
    \vspace{10mm} \subfigure{\epsfig{file=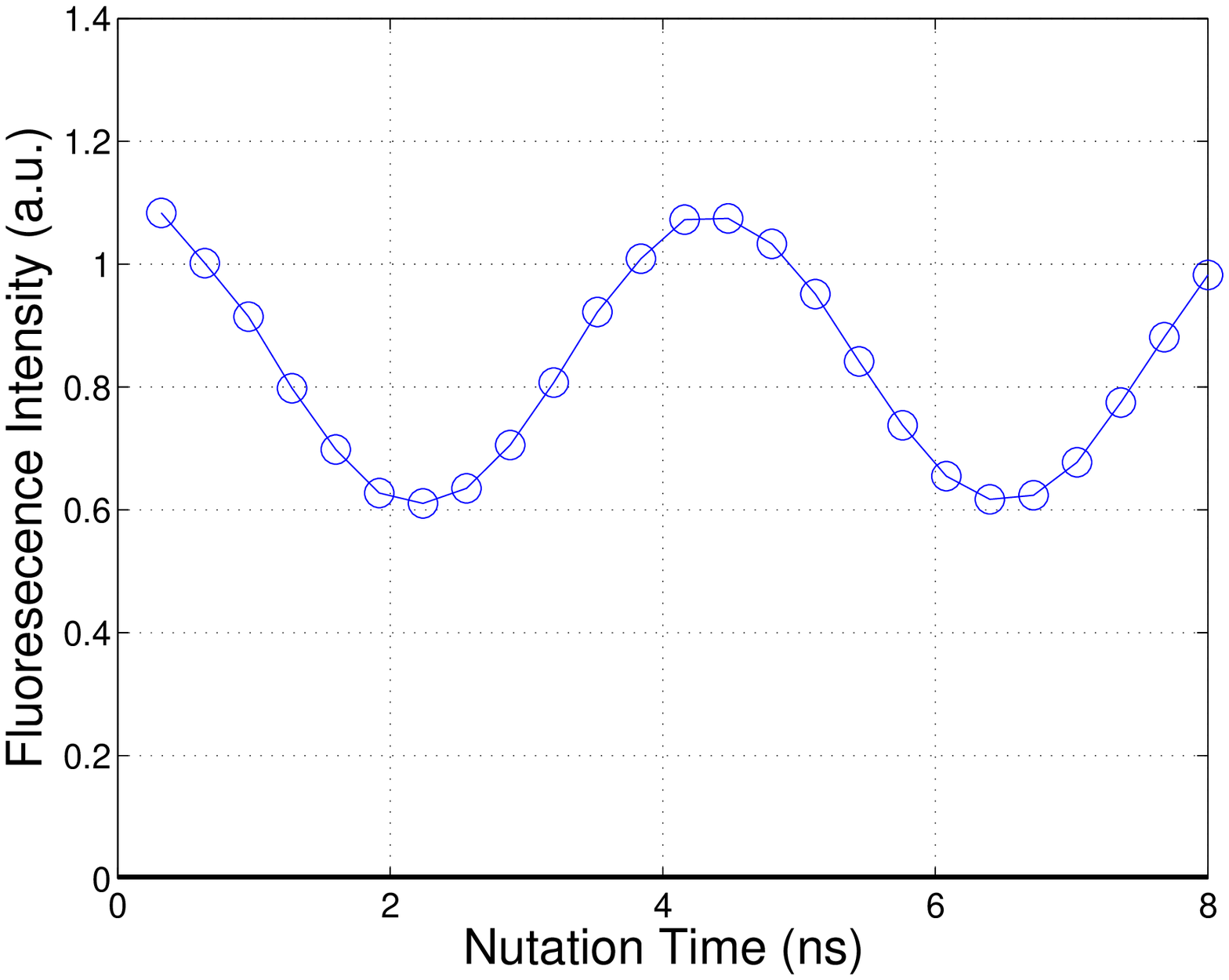, height=50mm}}{{(b)}}
  \caption{\label{triplet} (a) Energy level schematic of spin S=1 triplet in an applied magnetic
  field. Because of the axial symmetry of the N-V defect, the $m_S=\pm1$
  sublevels remain degenerate in zero field and are split in non-zero magnetic field. (b) State readout is via Rabi oscillations
   of fluorescence intensity. The circles represent experimentally measured data points.}
  \end{center}
\end{figure}

The energy level scheme of the N-V center is shown in figure \ref{levels}(b). It consists of a triplet ground state $^3A$ and a triplet excited
state $^3E$ with a metastable singlet state $^1A$. At zero magnetic field the ground triplet state energy levels are split into degenerate
$m_s=\pm1$ sublevels and the $m_s=0$ sublevel. This is due to the magnetic dipole-dipole interaction between the 2 unpaired electrons. This
dipole-dipole interaction can be described by a Hamiltonian term $H_D=\textbf{S}\cdot\overline{D}\cdot \textbf{S}$ where $\overline{D}$ is the
zero-field splitting (or fine structure) tensor and $\textbf{S}=(S_x,S_y,S_z)$. Using standard techniques \cite{epr} this dipolar term can be
rewritten as
\begin{equation} H_D = D\left[S^2_z-\frac{1}{3}S(S+1)\right]+E(S^2_x+S^2_y),\end{equation} where $D$ and $E$ are zero-field splitting parameters
which are dependent on the symmetry of the molecule. The axial symmetry of the N-V center means that $E=0$ MHz, while $D$ for this system has
previously been characterized \cite{vanoort,reddy} as $D=2880$ MHz. The end result is that, at zero magnetic field, the $m_s=\pm1$ sublevels are
degenerate and at an energy 2880 MHz greater than that of the $m_s=0$ sublevel. With an applied magnetic field aligned along the z axis of the
molecular system (i.e. $\vec{B}=(B_z,0,0)$) the total Hamiltonian, including zero-field splitting term, takes the simple form
\begin{equation} H_{Tot}=g\beta B_z S_z + D\left[S_z^2-\frac{1}{3}S(S+1)\right],
\end{equation}
where the gyromagnetic ratio, $g\beta$, is 2.8025 MHz/Gauss. This Hamiltonian leads to a $\ket{0} \leftrightarrow \ket{1}$ transition of energy
\begin{equation}
\Delta E = (2,880 + 2.8025 B_z) $MHz$ \label{transition}
\end{equation}

where $B_z$ is the applied magnetic field (figure \ref{triplet}(a)), measured in Gauss.

%
%
%

The fluorescence intensity corresponding to the $m_s=0$ transition between the ground $(^3A)$ and excited $(^3E)$ triplet states is strongest.
Optically detected magnetic resonance (ODMR) is used to perform state readout \cite{thir,fift,sixt}, i.e. the population of the $m_s=0$ sublevel
is identified by the amount of measured fluorescence (figure \ref{triplet}(b)). Manipulation of the qubit state is via standard electron spin
resonance (ESR) techniques, using microwave pulses resonant with the $\ket{0} \leftrightarrow \ket{1}$ transition \eref{transition}.

The spin longitudinal relaxation time, $T_1$, is on the order of milliseconds at room temperature \cite{Redman} (relaxation time of the order of
seconds is expected at low temperature). Transverse relaxation times (or decoherence times), $T_2$, of up to 60 microseconds have been reported
\cite{twtw}, for samples with low nitrogen concentration.

In the N-V center, initialization into the $m_s=0$ state is achieved by optical pumping. Optically induced spin polarization is thought to be
related to spin-selective intersystem crossing from the photoexcited $^3E$ triplet state to the metastable $^1A$ singlet state (Figure
\ref{levels}(b)) i.e there is an irreversible transition between $^3E \rightarrow {^1A}$. The spin polarization achieved by this pumping
corresponds to at least 70\% population in the $m_s=0$ sublevel at room temperature. Instead of a pure state, corresponding to 100\%
polarization, we have what is known as a pseudopure state which takes the general form
\begin{equation}
\rho_{pseudo}=\left(\frac{1-\alpha}{2^n}\right)\mathbb{I}+\alpha\dm{\psi}{\psi} \label{pseudo}
\end{equation}
where $n$ is the number of qubits and $0 < \alpha < 1$ is related to the amount of polarization.

In the case of the N-V center, the 70\% population of the $m_s=0$ sublevel, after optical pumping, corresponds to $ ${Tr}$\left(\rho_{pseudo}
\dm{0}{0}\right)=.7$ and so
\begin{equation}
\rho_{pseudo}=.6\left(\frac{\mathbb{I}}{2}\right)+.4\dm{\psi}{\psi}.
\end{equation}

It is hoped that, in future experiments, the use of projective readouts, which are possible at low temperatures, will lead to increased
polarizations.

A controlled two-qubit quantum gate in which the vacancy electron spin is hyperfine coupled to a nearby C${}^{13}$ nucleus, has recently been
performed using this system \cite{twon}.


\section{Standard QPT}
\subsection{\textit{QPT Technique}}
We briefly review the standard QPT technique:
\begin{description}
\item{$\bullet$ Prepare a complete basis of input states $\rho_1\dots\rho_{d^2}$} (e.g $\rho_j=|\psi_j\rangle
\langle\psi_j|$,$|\psi_j\rangle=\{|0\rangle,|1\rangle,\frac{1}{\sqrt{2}} (|0\rangle+|1\rangle),\frac{1}{\sqrt{2}}(|0\rangle+i|1\rangle)\}$, for
a single qubit) \item{$\bullet$ Apply the unknown process \E to each member $\rho_j$}
\item{$\bullet$ Reconstruct the output states $\mathcal{E}(\rho_j)$ using quantum state tomography}
\end{description}
While each of the reconstructed $\mathcal{E}(\rho_j)$ will be physically valid density matrices (by appropriate use of quantum state tomography)
it is possible that the process \E describes is unphysical e.g. the $\mathcal{E}(\rho_j)$ are mutually inconsistent or \E is not completely
positive. \textbf{Experimental noise and a finite number of measurements to determine expectation values can result in a reconstructed \E which
is not physically valid}.

If it is possible to write a process in the Kraus representation then we can be sure it represents a physically valid process (one that is
described by a completely positive map). We can specify an unknown process \E by experimentally determining the Kraus operators \cite{quantop}
$\{E_i\}$ which describe it
\begin{equation}
\Er=\sum_{i=1}^{d^2}E_i \rho E_i^{\dag}. \label{krausy}
\end{equation}
or, using a fixed (complete) basis of operators $\{A_i\}$
\begin{equation}
\mathcal{E}(\rho)=\sum_{m,n=1}^{d^2} \chi_{mn}\hspace{3pt} { A}_m \rho {A}^\dag_n. \label{aman}
\end{equation}

$\chi_{mn}$ is a matrix of coefficients which completely describes the process $\mathcal{E}$ and is positive Hermitean by construction.   The
trace preserving constraint $\sum_i E_i^{\dag} E_i=\mathbb{I}$ becomes $\sum_{m,n} \chi_{mn} A_n^{\dag}A_m=\mathbb{I}$.

\begin{description}
\item{$\bullet$ From the measurement
results $\{\mathcal{E}(\rho_1)\dots\mathcal{E}(\rho_{d^2})\}$, $\lambda_{jk}$ can be determined, given the relation ${\cal
E}(\rho_j)=\lambda_{jk}\rho_k$.}
\item{$\bullet$ In order to determine $\chi$ from the matrix $\lambda$ one operates on $\lambda$ with the pseudoinverse
of $\beta$, where $\beta$ is derived theoretically from the relation ${\hat A}_m \rho_j {\hat A}^\dag_n = \beta^{mn}_{jk} \rho_k$.}
\item{$\bullet$ Using this last relation and \eref{aman} we can see $\lambda_{ij} =\sum_{mn} \beta^{mn}_{ij} \chi_{mn}$ and so inverting $\beta$
gives us $\chi$, as required.}

To derive $\{E_i\}$ from $\chi$ we first diagonalize $\chi$ with a unitary $U^{\dag}$
\begin{equation}
\chi_{mn}=\sum_{k,l} U_{mx}d_k \delta_{kl} U^{*}_{nl}
\end{equation}
where $d_i$ are the eigenvalues of $\chi$. The Kraus operators can then be obtained by
\begin{equation}
E_i=\sqrt{d_i}\sum_j U_{ji}A_j
\end{equation}
\end{description}

Note that this procedure only works if $d_i\geq 0$ which follows from the positivity (semidefinite) of $\chi$. It is this property which we use
to check the physicality of a process $\mathcal{E}$; if the \x matrix reconstructed from experimental data has negative eigenvalue(s) then this
indicates that noise and/or finitely sampled expectation values has caused the output data to infer an unphysical process. To overcome this
problem, a physical matrix \xt is found which is as close as possible to the original \x in some sense. Specifically, we used a technique
analogous to MLE for state estimation \cite{jezek,cnotqpt} by minimising a deviation function, $\Delta(t)$, incorporating a general
parameterization for a positive \xt:
\begin{equation}
\Delta(t)=\sum_{m,n=1}^{d^2}\left|\xxt_{mn}(t)-\xx_{mn}\right|^2 + \lambda\left|\sum_{m,n=1}^{d^2}\xxt_{mn}(t)A_n^{\dag}A_m-\mathbb{I}\right|^2
\label{min}
\end{equation}
where $\lambda$ is a Lagrange multiplier (which ensures a trace-preserving process) and
\begin{equation}
\xt=T^{\dag}(t)T(t)
\end{equation}
where $T(t)$ is a $d^2 \times d^2$ complex, lower triangular matrix with $d^4$ real parameters, $t(i)$. In the case of a single qubit we have:
\begin{equation}
T(t)=\left(%
\begin{array}{cccc}
  t(1) & 0 & 0 & 0 \\
  t(5)+it(6) & t(2) & 0 & 0 \\
  t(11)+it(12) & t(7)+it(8) & t(3) & 0 \\
  t(15)+it(16) & t(13)+it(14) & t(9)+it(10) & t(4) \\
\end{array}%
\right).\label{Tp}
\end{equation}
The algorithm used to minimise $\Delta(t)$ in \eref{min} was the Nelder-Mead simplex algorithm as implemented in Matlab \textregistered.


\subsection{\textit{Avoiding Local Minima}}
As is often the case with numerical minimization problems, \eref{min} typically contains numerous local minima. Our preliminary investigations
suggested that, given a random initial point in parameter space, the algorithm would often fail to find the global minimum. As such, a good
starting point, $t^{\bigstar}(i)$, in \eref{Tp} was deemed necessary if the results were to be meaningful. In order to obtain $t^{\bigstar}(i)$
we used a technique based on principal component analysis \cite{pca}. If the experimentally determined matrix \x is not positive semidefinite,
one can ``filter" it by setting any (presumably quite small) negative eigenvalues to zero. Specifically, if we decompose \x as
\begin{equation}
\x=U D U^{\dag},
\end{equation}
where $D$ is a diagonal matrix containing the eigenvalues of \x, then we can construct a similar but positive matrix $\chi^{\bigstar}$ via
\begin{equation}
\chi^{\bigstar}=U D^{\bigstar} U^{\dag},
\end{equation}
where $D^{\bigstar}$ is identical to $D$ except that any negative eigenvalues have been set equal to zero. From $\chi^{\bigstar}$ we can extract
good initial parameters $t^{\bigstar}(i)$ by performing a Cholesky decomposition (e.g. using the in-built ``Chol" function in Matlab
\textregistered).


\subsection{\textit{Process Visualisation}}
Any one qubit state can be parameterized by a so-called Bloch vector $\vec{r}$. Equivalently, one can explicitly include the unit coefficient of
the identity basis component and parameterise the qubit state by a 4-vector $(1,r_x,r_y,r_z)^T$:
\begin{equation}
\rho=\frac{1}{2}\left(\mathbb{I}+\vec{r}\cdot\vec{\sigma}\right)\longleftrightarrow\frac{1}{2}\left(%
\begin{array}{c}
  1 \\
  r_x \\
  r_y \\
  r_z \\
\end{array}%
\right)
\end{equation}
where $r_x^2+r_y^2+r_z^2\leq1$ \cite{incomplete}. In this basis any linear trace-preserving evolution \E takes the affine form
\begin{equation}
\mathcal{E}_{\mathcal{A}}=\left(%
\begin{array}{cc}
  1 & 0 \\
  \vec{t} & E \\
\end{array}%
\right)=\left(%
\begin{array}{cccc}
  1 & 0 & 0 & 0 \\
  t_x & E_{xx} & E_{yx} & E_{zx} \\
  t_y & E_{xy} & E_{yy} & E_{zy} \\
  t_z & E_{xz} & E_{yz} & E_{zz} \\
\end{array}%
\right),
\end{equation}
and the process applied to an arbitrary input state is an affine map on $\vec{r}$, (although it remains a linear map on $\rho$):
\begin{equation}
\left(%
\begin{array}{cc}
  1 & 0 \\
  \vec{t} & E \\
\end{array}%
\right)\left(%
\begin{array}{c}
  1 \\
  \vec{r} \\
\end{array}%
\right)=\left(%
\begin{array}{c}
  1 \\
  E\vec{r}+\vec{t} \\
\end{array}%
\right)=\left(%
\begin{array}{c}
  1 \\
  \vec{r\prime} \\
\end{array}%
\right),
\end{equation}
or, more explicitly,
\begin{equation}
\mathcal{E}_{\mathcal{A}}(\vec{r})=\vec{r\prime}=E\vec{r}+\vec{t}.
\end{equation}

 \textbf{Identifying the surface of the unit Bloch sphere as the set of all possible input states, we can visualize a process \E by its
action on the sphere}. The $3 \times 3$ matrix $E$ is responsible for deformation and rotation of the Bloch sphere, while $\vec{t}$ denotes
displacement from $\vec{r}=(0,0,0)$.

We can gain crude but immediate insight into the nature of a single qubit process by using this visualization technique. Bloch vectors with
$|\vec{r}| > 1$ do not correspond to any valid density matrix. Hence, when a process is depicted this way, any protrusions of the output
ellipsoid outside the unit Bloch sphere mean that the process is not trace preserving.

Complete positivity also places constraints on physically allowable output ellipsoids. For a more detailed analysis of this problem we refer to
\cite{ruskai,oi,akio}. The crucial point to note is that seemingly innocuous ellipsoids (i.e. contained within the unit Bloch sphere and not
overly deformed) can represent a process which is not physically valid. The variety of allowable ellipsoids is particularly reduced when the
translation vector $\vec{t}$ in $\mathcal{E}_{\mathcal{A}}$ is non-zero.


\section{Experimental Method}
The QPT experiment was performed at room temperature using diamond nanocrystals obtained from type Ib synthetic diamond. Diamond nanocrystals
were spin coated on a glass substrate, and single nanocrystals were observed with a homebuilt sample-scanning confocal microscope. In order to
ensure the presence of a single defect in the laser focus, the second-order coherence was measured using Hanbury-Brown and Twiss interferometer
and the contrast of the antibunching depth was determined.  In order to perform many repetitions (to obtain good expectation values) in a
reasonable amount of time, a sample with a relatively short coherence time ($2\mu$s) was chosen. Microwaves were coupled to the sample by a ESR
microresonator connected to a 40W travelling wave tube amplifier. A magnetic field ($\vec{B}=(0,0,B_z)$, $B_z=200$ Gauss) was applied to the
defect in order to remove degeneracy of the $\ket{\pm1}$ energy levels.

Rabi oscillations of fluoresence intensity were used to obtain expectation values. A reference nutation was initially taken in order to
normalize the measurement nutations (i.e. set 0,1 levels for expectation values for the pseudopure initial state). In addition this reference
oscillation was used to derive the microwave pulse time required to perform $\frac{\pi}{2}$ and $\pi$ rotations of the Bloch vector. A complete
basis of four input states was then prepared using microwave pulses resonant with qubit transitions. The $\rho_{0}$ state  was obtained directly
by optical pumping and three remaining input states $\rho_{1}$, $\rho_{2}$ and $\rho_{3}$ were obtained by application of suitable $\pi$ or
${\frac{\pi}{2}}$ pulses. Each of these states was left to decohere for a series of time intervals $\tau$. As a last step, measurements of the
diagonal and off-diagonal elements of the density matrix were performed.

Estimates of the density matrix elements were extracted from experimental data (Rabi oscillations) using the maximum entropy (MaxEnt) technique
\cite{MaxEnt}. This method returns a physically valid density matrix which satisfies, as closely as possible, the expectation values of measured
observables. In cases where only incomplete knowledge of the output state is known, an additional constraint is used; the reconstructed state
must also have the maximum allowable von Neumann entropy.


\section{Quantum process tomography experiment}
In summary the QPT experiment was performed as follows:

\begin{description}
\item $\bullet$ The N-V center was optically pumped to the initial pseudopure state
\begin{equation}
\varrho_{pseudo}=.6\left(\frac{\mathbb{I}}{2}\right)+.4\dm{0}{0}
\end{equation}
Hereafter we neglect the identity component and treat this initial state as $\rho=\dm{0}{0}$.

\item $\bullet$ Following this, each of the following input states $\rho_1\dots\rho_{d^2}$ ($\rho_j=|\psi_j\rangle
\langle\psi_j|$,$|\psi_j\rangle=\{|0\rangle,|1\rangle,\frac{1}{\sqrt{2}} (|0\rangle+|1\rangle),\frac{1}{\sqrt{2}}(|0\rangle+i|1\rangle)\}$ was
created by application of suitable $\pi$ or $\frac{\pi}{2}$ pulses.


\item $\bullet$ Each input state was then left to decohere, solely by interaction with its surroundings (i.e. no microwave (ESR) or optical
(ODMR) radiation was applied to the system), for a decoherence time, $\tau$.

\item $\bullet$ After $\tau$ had elapsed, expectation values for $\hat{\sigma_x}$,$\hat{\sigma_y}$ and $\hat{\sigma_z}$ were obtained by ODMR.

\item $\bullet$ Using $<\hat{\sigma_x}>$,$<\hat{\sigma_y}>$ and $<\hat{\sigma_z}>$ the final (output) states $\mathcal{E}(\rho_i)$, after the decoherence, were reconstructed via the MaxEnt
technique.

\item $\bullet$ Using these output states, the process for each time period was reconstructed using the technique described above i.e. minimization of
\eref{min}
\end{description}

\begin{figure}[!ht]
    \begin{center}
    \framebox{
    \subfigure{\epsfig{file=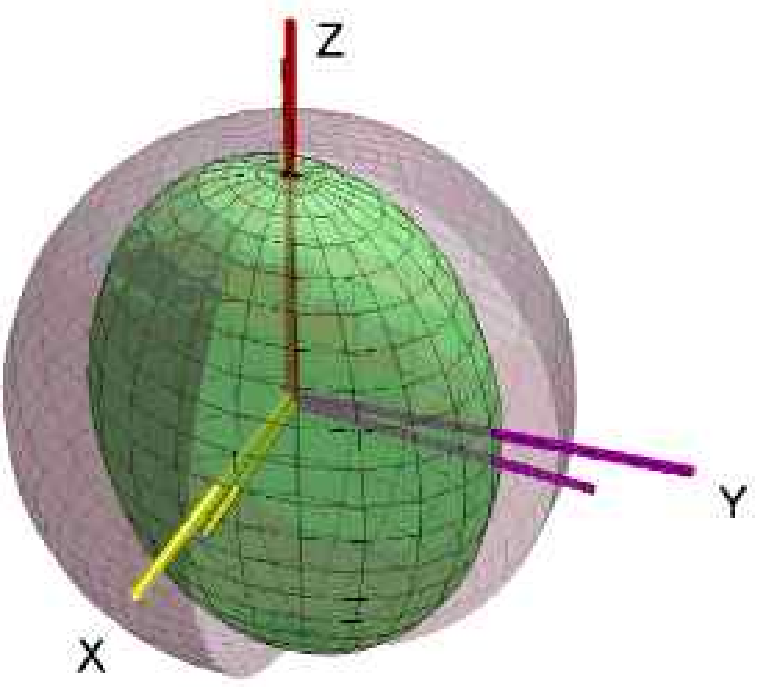,width=60mm}}{{a(i)}}
        \subfigure{\epsfig{file=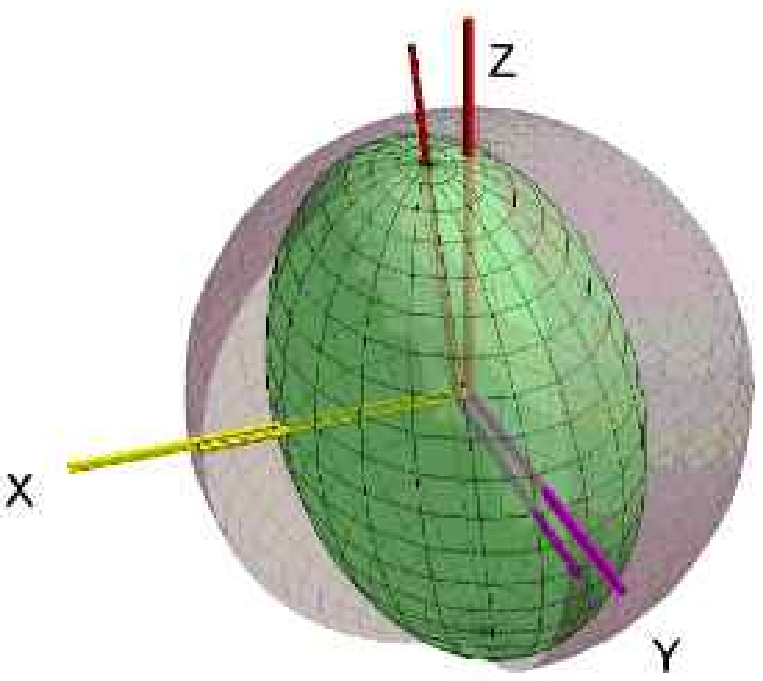,width=60mm}}{{a(ii)}}}
    \framebox{\subfigure{\epsfig{file=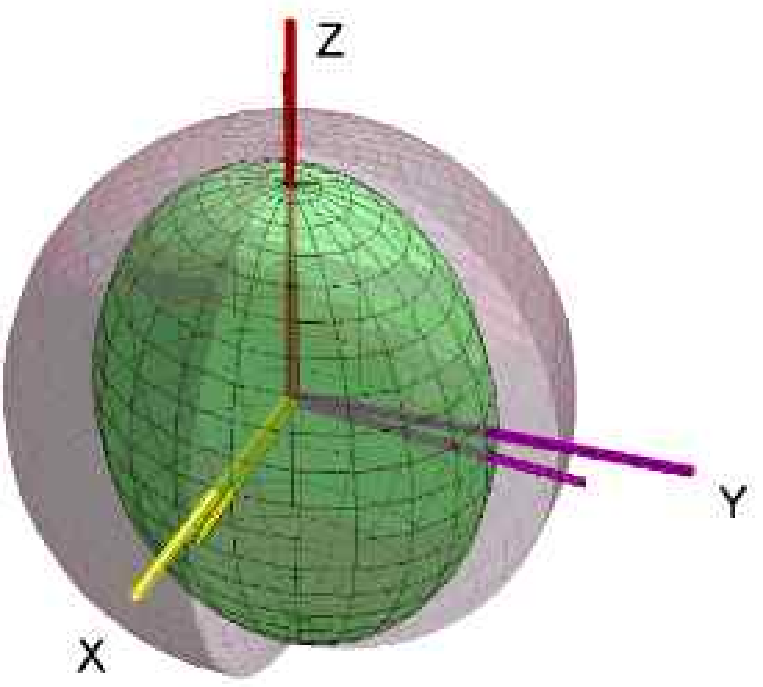,width=60mm}}{{b(i)}}
    \subfigure{\epsfig{file=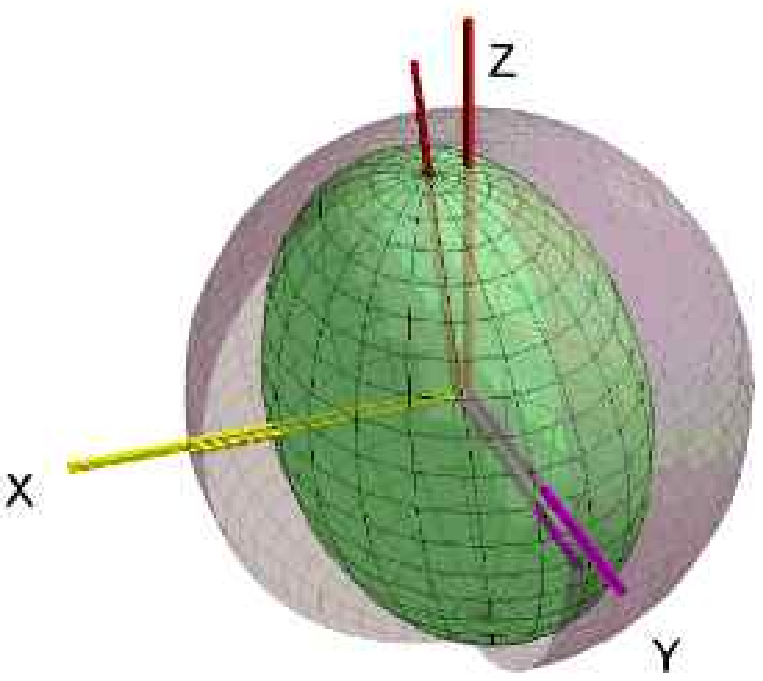,width=60mm}}{{\hspace{1mm}b(ii)}}}
\caption{20ns results: (a) Process from experimental data, $\chi$. (b) Process obtained from physically valid $\tilde{\chi}$. }
\end{center}
\begin{center}\protect\begin{tabular}[!ht]{|c|c|c|}
    \hline
&\  \ \ \ Process Matrix $\mathcal{E}_{\mathcal{A}}$ \  \ \\
\hline
 Experimental   &\quad $\left(%
\begin{array}{cccc}
    1.0000   &      0   &      0   &      0\\
    0.0626  &  0.6552  & -0.0225 &  -0.1198\\
    0.0448  &  0.0287  &  0.7309 &  -0.0226\\
    0.0138  & -0.0143  &  0.0878 &   0.9843
\end{array}%
\right)$\ \ \quad\\
    \hline
 Reconstructed    & $\left(%
\begin{array}{cccc}
    1.0000  &       0  &       0   &      0\\
    0.0532  &  0.6798  & -0.0312  & -0.1093\\
    0.0420   & 0.0206   & 0.7051  & -0.0227\\
    0.0070   & 0.0001  &  0.0916  &  0.9410
\end{array}%
\right)$  \\
    \hline
  \end{tabular}\end{center}
\label{fig20}
\end{figure}

\begin{figure}[!ht]
    \begin{center}
    \framebox{
    \subfigure{\epsfig{file=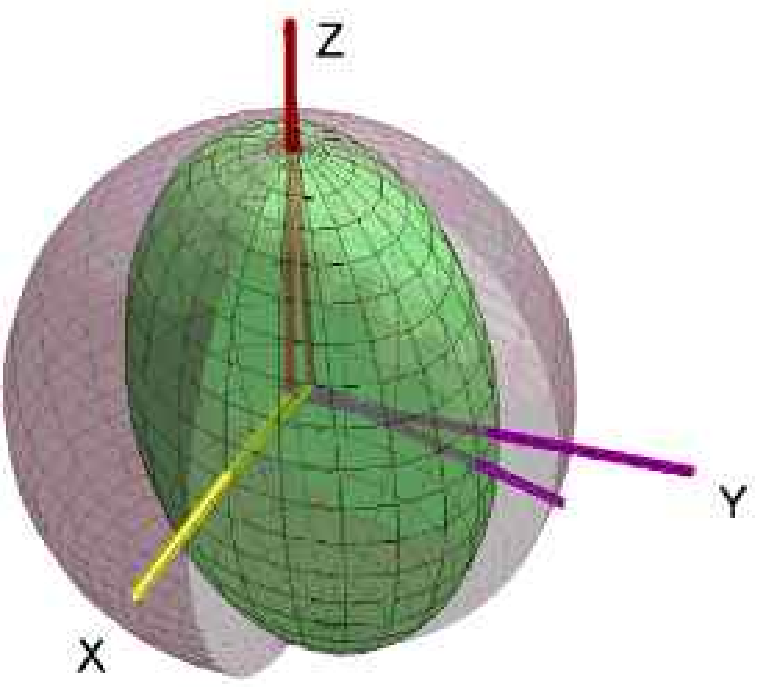,width=60mm}}{{a(i)}}
        \subfigure{\epsfig{file=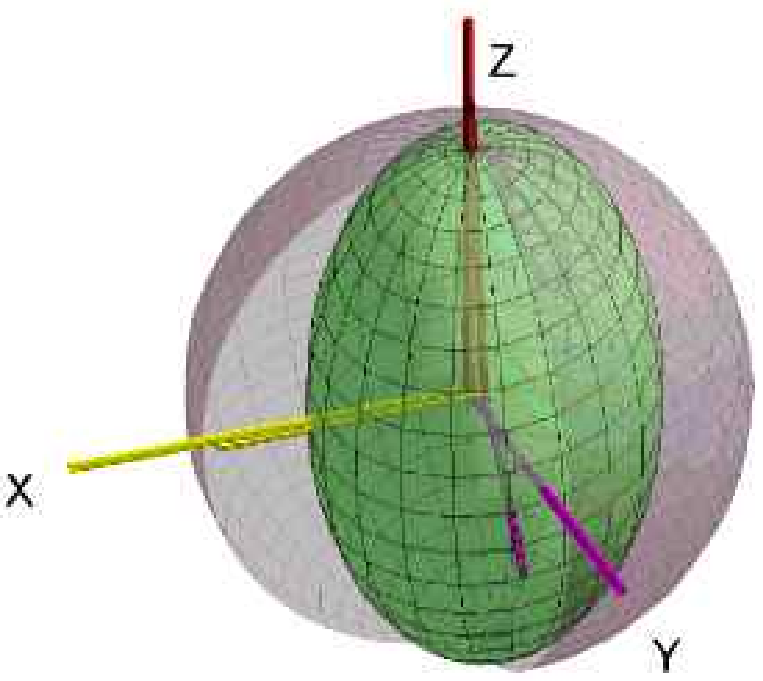,width=60mm}}{{a(ii)}}}
    \framebox{\subfigure{\epsfig{file=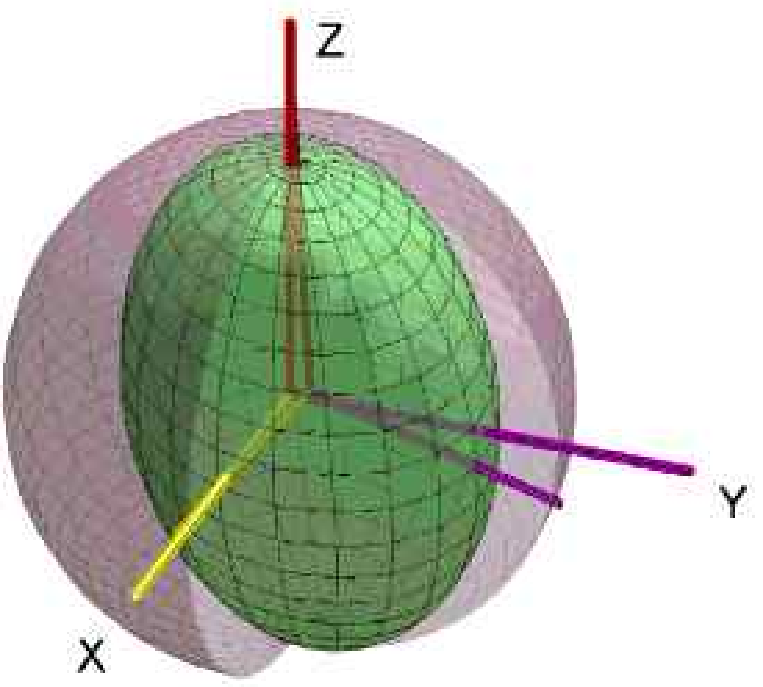,width=60mm}}{{b(i)}}
    \subfigure{\epsfig{file=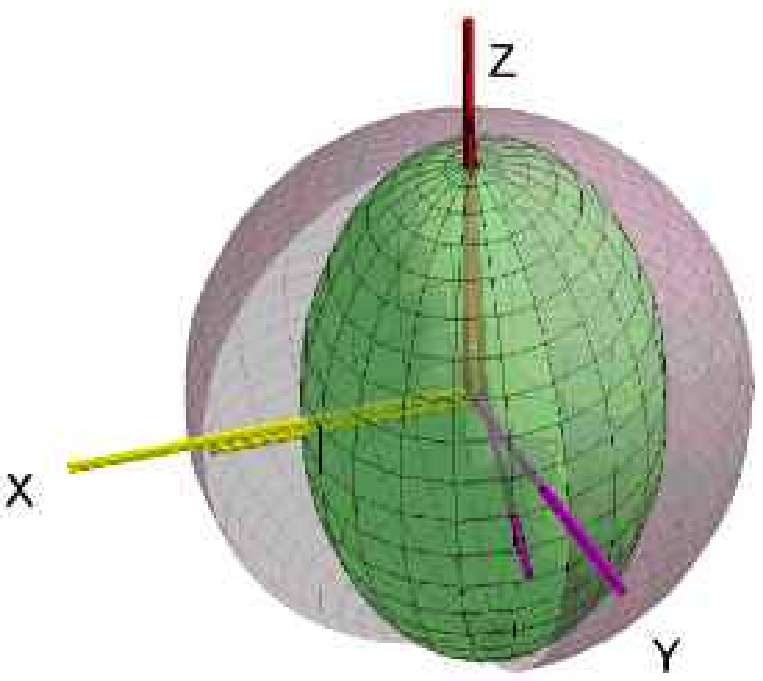,width=60mm}}{{\hspace{1mm}b(ii)}}}
\caption{40ns results: (a) Process from experimental data, $\chi$. (b) Process obtained from physically valid $\tilde{\chi}$. }
\end{center}
\begin{center}\protect\begin{tabular}[!ht]{|c|c|c|}
    \hline
&\  \ \ \ Process Matrix $\mathcal{E}_{\mathcal{A}}$ \  \ \\
\hline
 Experimental   &\quad $\left(%
\begin{array}{cccc}
    1.0000  &       0   &      0   &      0\\
   -0.0344  &  0.6378 &   0.1660 &  -0.0261\\
    0.0683  & -0.0728 &   0.6872 &   0.0499\\
    0.0277  &  0.0536 &   0.0827 &   0.9614
\end{array}%
\right)$\ \ \quad\\
    \hline
 Reconstructed    &\quad $\left(%
\begin{array}{cccc}
    1.0000  &       0  &       0  &       0\\
   -0.0307  &  0.6477  &  0.1443  & -0.0266\\
    0.0638  & -0.0944  &  0.6773  &  0.0407\\
   -0.0004  &  0.0532  &  0.0824  &  0.9320
\end{array}%
\right)$\ \ \quad \\
    \hline
  \end{tabular}\end{center}
\label{fig40}
\end{figure}

\begin{figure}[!ht]
    \begin{center}
    \framebox{
    \subfigure{\epsfig{file=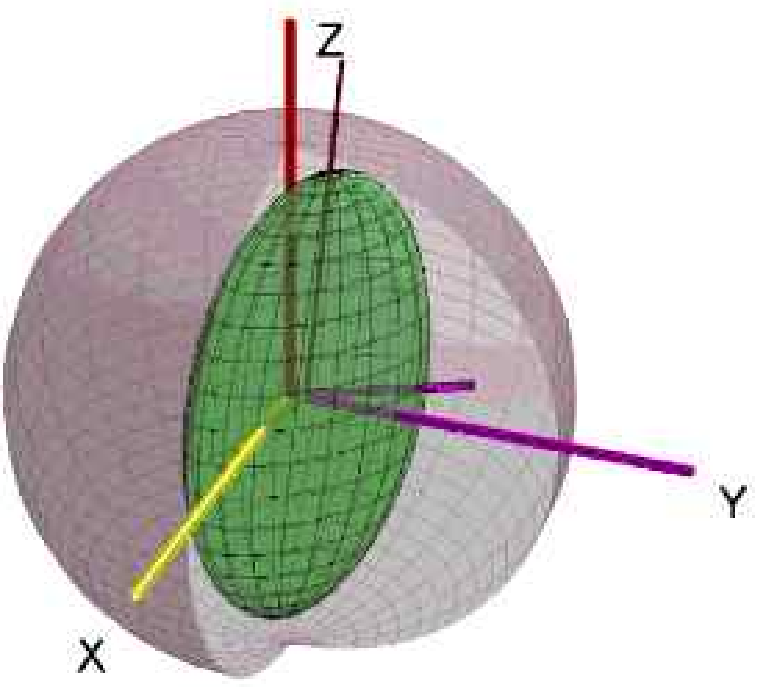,width=60mm}}{{a(i)}}
        \subfigure{\epsfig{file=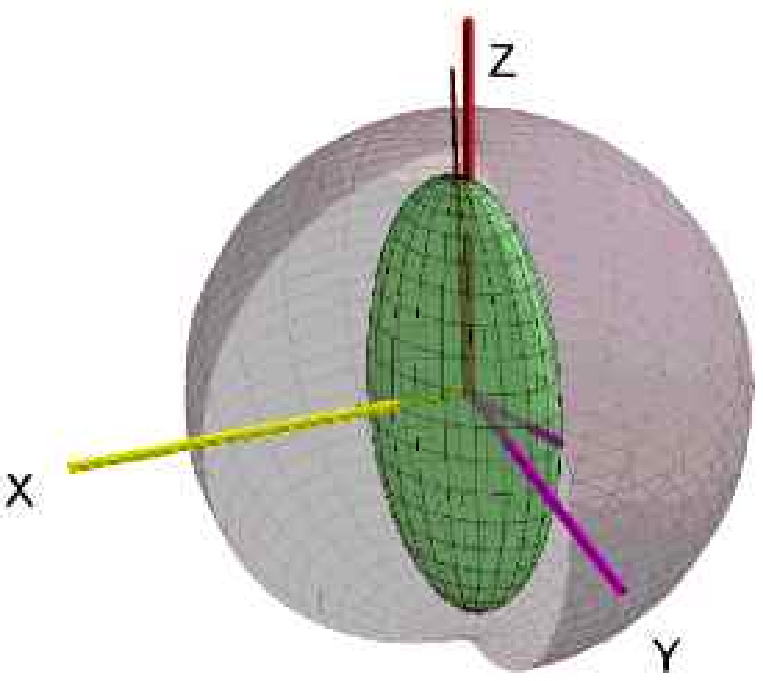,width=60mm}}{{a(ii)}}}
    \framebox{\subfigure{\epsfig{file=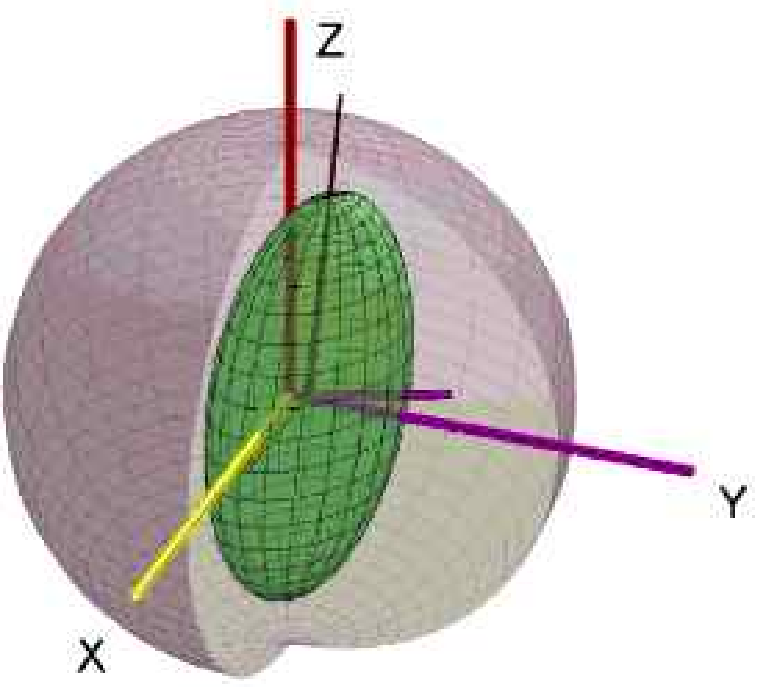,width=60mm}}{{b(i)}}
    \subfigure{\epsfig{file=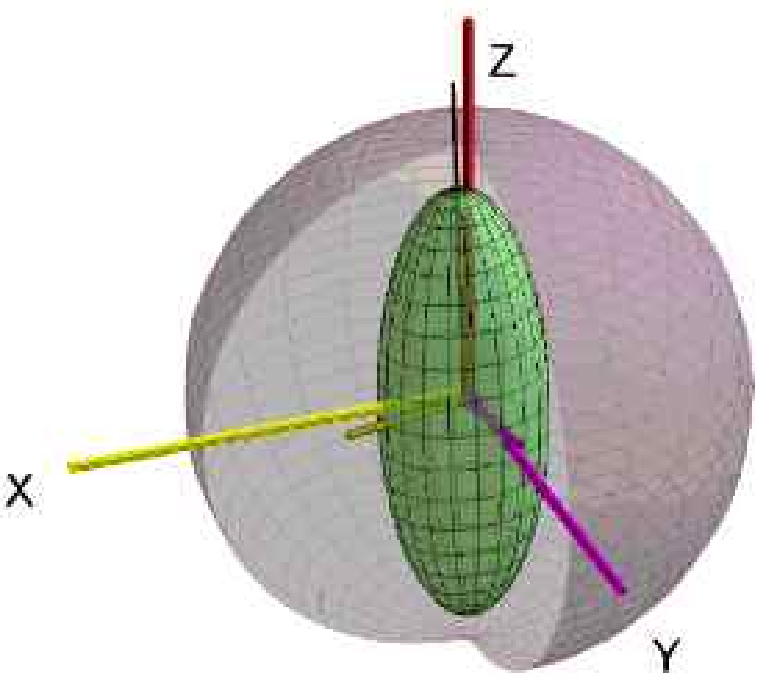,width=60mm}}{{\hspace{1mm}b(ii)}}}
\caption{80ns results: (a) Process from experimental data, $\chi$. (b) Process obtained from physically valid $\tilde{\chi}$. }
\end{center}
\begin{center}\protect\begin{tabular}[!ht]{|c|c|c|}
    \hline
&\  \ \ \ Process Matrix $\mathcal{E}_{\mathcal{A}}$ \  \ \\
\hline
 Experimental   &\quad $\left(%
\begin{array}{cccc}
    1.0000  &       0  &       0 &        0\\
    0.0442  &  0.2359  & -0.1001 &  -0.0757\\
    0.0791  & -0.0947  &  0.3770 &  -0.1163\\
    0.0434  &  0.0487  & -0.0461  &  0.9554
\end{array}%
\right)$\ \ \quad\\
    \hline
 Reconstructed    &\quad $\left(%
\begin{array}{cccc}
    1.0000  &       0  &       0   &      0\\
    0.0503  &  0.2726  & -0.0290  & -0.0752\\
    0.0834  & -0.0357  &  0.3291  & -0.1038\\
    0.0302  &  0.0457  & -0.0495  &  0.8984
\end{array}%
\right)$\ \ \quad \\
    \hline
  \end{tabular}\end{center}
\label{fig80}
\end{figure}

The processes, both from raw data and reconstructed (i.e. completley positive), are depicted in figures \ref{fig20} to \ref{fig80}.


\subsection{\textit{Quantification of Unphysicality}}

The Jamiolkowski isomorphism \cite{jamiolkowski} maps a quantum operation $\mathcal{E}$, acting on $\mathcal{H}_{d}$, to a quantum state \re \
in $\mathcal{H}_{d^2}$ via
\begin{equation}
\re\equiv\left[\mathbb{I}_{d}\otimes\mathcal{E}\right]\left(\dm{\Phi}{\Phi}\right) \label{jamiso}
\end{equation}
where $\dm{\Phi}{\Phi}$ is a projector on to a maximally entangled state in $\mathcal{H}_{d^2}$ i.e.
\begin{equation}
\ket{\Phi}=\sum_j \frac{\ket{j} \ket{j}}{\sqrt{d}} \label{ent}
\end{equation}
where $\left\{\ket{j}\right\}$ is some orthonormal basis set.

%

When the process $\mathcal{E}$ is physical, one then obtains a physically valid $\rho_{\mathcal{E}}$ which can be compared to the ideal process
(converted to $\rho_{id}$) using distance measures on quantum states \cite{distance}. The \textbf{trace distance} between density matrices
$\rho_{id}$ and $\rho_{\mathcal{E}}$ is
\begin{equation}D(\rho_{id},\rho_{\mathcal{E}})\equiv \frac{1}{2} tr \mid \rho_{id}-\rho_{\mathcal{E}}\mid,\end{equation} where \begin{equation}\mid X\mid \equiv \sqrt{X^\dag
X}.\end{equation} Similarly one can define the {\bf Fidelity}: \begin{equation}F(\rho_{id},\rho_{\mathcal{E}})\equiv tr
\left(\sqrt{\sqrt{\rho_{id}}\ \rho_{\mathcal{E}} \sqrt{\rho_{id}}}\right)^2.\end{equation} Using this definition we define the \textbf{Bures
Metric}
\begin{equation}
B(\rho_{id},\rho_{\mathcal{E}})\equiv \sqrt{2-2\sqrt{F(\rho_{id},\rho_{\mathcal{E}})}}\end{equation} and the \textbf{C Metric}
\begin{equation}C(\rho_{id},\rho_{\mathcal{E}}) \equiv \sqrt{1-{F(\rho_{id},\rho_{\mathcal{E}})}}.
\end{equation}
An unphysical process, however, can lead to an unphysical $\rho_{\mathcal{E}}$, possibly resulting in a process fidelity which is greater than
one. The application of the preceding fidelity-based distance measures can, therefore, produce nonsensical results. In such cases it is
necessary to use other techniques in order to estimate the disparity between $\chi$ and $\tilde{\chi}$ for example. If one defines $X
=\chi-\tilde{\chi}$, then possible measures are the matrix p-norms ($p=1,2,\infty$) of $X$ and the Frobenius norm of $X$ ($\|X\|_{Fro}$) as well
as the trace distance ($D_{pro}$).

As stated above, these quantities gives a measure of how well $\tilde{\chi}$ describes the experimental results. The difference matrix $X$
varies depending on the basis of operators chosen for $\chi$. We propose that if a standard basis is chosen then the norms of $X$, defined
above, provide a method of ``benchmarking" the quality of all such quantum process tomography experiments, regardless of implementation. Table
\ref{discrep} describes the discrepancies between experimental and reconstructed processes. Here we have chosen the normal basis $\{A_k\} (
A_{di+j+1}=\dm{i}{j})$ with which to construct $\chi$ and \xt. This particular choice of basis means that \x is equal to the Choi matrix
$\mathcal{C}$ \cite{Choi,havel}
\begin{equation}
\x=\mathcal{C}=\sum_{i,j=0}^{d-1} \mathcal{E}\left(\dm{i}{j}\right) \otimes \dm{i}{j},
\end{equation}
and proportional to the characteristic state $\re$\ for the process:
\begin{equation}
\x=d\re =d \left(\mathcal{E}\otimes \mathbb{I}_d\right)\dm{\Phi}{\Phi}.
\end{equation}

\begin{table}[!ht]
\begin{center}
\begin{tabular}{|c|c|c|c|c|}
  \hline
  Decoherence & & & & \\
  time&  $\|X\|_{p=1}$ & $\|X\|_{p=2}$ & $\|X\|_{Fro}$ & $D_{pro}$ \\
  \hline
 20ns &  0.0660 &   0.0525  &  0.0636  &  0.0262 \\
 \hline  40ns & 0.0581  &  0.0427  &  0.0529  &  0.0213 \\
\hline  80ns & 0.1141  &  0.1075   & 0.1276   & 0.0538 \\
  \hline
\end{tabular}
\caption{\label{discrep}The disparity between the estimated process from the experimental measurements ($\chi$) and the ``nearest'' physically
valid process ($\tilde{\chi}$). $X$ is defined as $X =\chi-\tilde{\chi}$. }
  \end{center}
\end{table}


\section{Markovian Process Tomography}

Standard QPT utilises a ``black box" approach to studying dynamics. It predicts the resultant output states given arbitrary initial states, but
fails to describe the path taken through state space, over time, from initial to final state. If we are prepared to make certain assumptions
about the relationship between system and environment, the Born and Markov approximations, then we can construct a Markovian quantum master
equation which describes the time evolution of the system over the time studied. In the area of quantum information processing, understanding
open system dynamics is important in studying quantum noise processes, designing error correcting codes \cite{GKtheorem} and locating
decoherence free subspaces \cite{dfs}. We discuss a method of experimentally reconstructing a master equation by using estimates of the state of
the system at a number of different timepoints. This particular technique was developed and implemented by Boulant, Havel, Pravia and Cory
\cite{havel}, in the context of liquid NMR quantum computation. The method is different to standard QPT techniques insofar as it assumes
Markovian dynamics and, consequentially, the reconstructed equation has separate terms which describe the unitary and non-unitary aspects of the
evolution. We then proceed to apply this technique to the N-V center, in an effort to better identify the decoherence processes it suffers. This
is the first time such an analysis has been performed on a single solid state qubit.

By invoking the Born and Markov approximations one may derive the so-called \textbf{Lindblad Master Equation} \cite{lindbladpap}, commonly
expressed in a number of different but equivalent ways, including:
\begin{eqnarray}
\frac{\partial \rho}{\partial t}&=&-i\left[H,\rho\right]+\frac{1}{2}\sum_{k=1}^{d^2-1}\left(2L_k\rho L_k^{\dag}-L_k^{\dag}L_k\rho-\rho
L_k^{\dag}L_k\right),\\
\frac{\partial \rho}{\partial t}&=&-i\left[H,\rho\right]+\frac{1}{2}\sum_{k=1}^{d^2-1}\left(\left[L_k\rho,L_k^{\dag}\right]+\left[L_k,\rho
L_k^{\dag}\right]\right).\label{lindbladb} \\
\frac{\partial \rho}{\partial t}&=&\hh{\mathcal{L}}\left[\rho\right]
\end{eqnarray}

The Lindblad master equation \eref{lindbladb} is the diagonal form of the GKS master equation derived by Gorini, Kossakowski and Sudarshan
\cite{gorini}:
\begin{equation}
\frac{\partial \rho}{\partial t}=-i\left[H,\rho\right]+\frac{1}{2}\sum_{\alpha,\beta=1}^{d^2-1} a_{\alpha
\beta}\left(\left[F_{\alpha}\rho,F_{\beta}^{\dag}\right]+\left[F_{\alpha},\rho F_{\beta}^{\dag}\right]\right)\label{eqgorini}
\end{equation}
where $a_{\alpha \beta}$ is a $(d^2-1) \times (d^2-1)$ Hermitean matrix and $\left\{F_{\alpha}\right\}$ is a linear basis of \emph{traceless}
operators on density matrices. If the matrix of coefficients $a_{\alpha \beta}$ (sometimes called the GKS matrix \cite{simulation}) is positive,
then the process described by \eref{eqgorini} is completely positive. Diagonalising $a_{\alpha \beta}$ in \eref{eqgorini} leads to the Linblad
form \eref{lindbladb}.

We can clearly see one advantage gained by making the Born-Markov approximations - the resulting master equation can be separated into
components which cause unitary and non-unitary evolution of the system:
\begin{eqnarray}
\rm{Unitary \ Evolution:} &\qquad -i\left[H,\rho\right],\label{unpart}\\
{\rm Non-Unitary \ Evolution:} &\qquad \frac{1}{2}\sum_{k=1}^{d^2-1}\left(\left[L_k\rho,L_k^{\dag}\right]+\left[L_k,\rho
L_k^{\dag}\right]\right).
\end{eqnarray}
The operators $L_k$ are \textbf{Lindblad operators} and they describe the decoherence processes.

We now describe a technique for estimating the generator $\hh{\mathcal{L}}$ of a Markovian process. We transform to a Liouville space basis, as
in \cite{havel}, where density matrices become column vectors (denoted \ur), and dynamical maps become $d^2 \times d^2$ matrix superoperators
(denoted by \hh{}\ ):
\begin{equation}
\frac{\partial \ur}{\partial t}=\hh{\mathcal{L}}\ur.
\end{equation}
The matrix superpropagator $\hh{\mathcal{P}}(t)$ is then defined by exponentiation:
\begin{equation}
\urt{t}=e^{\hh{\mathcal{L}}t}\urt{0}=\hh{\mathcal{P}}(t)\urt{0}.
\end{equation}
Henceforth we adopt a similar notation to \cite{havel}:
\begin{eqnarray}
\frac{\partial \rho}{\partial t}&=&-i\left[H,\rho\right]+\frac{1}{2}\sum_{k=1}^{d^2-1}\left(\left[L_k\rho,L_k^{\dag}\right]+\left[L_k,\rho
L_k^{\dag}\right]\right)\\ \nonumber
\\ \nonumber
{\rm becomes} \\ \nonumber
\\
\frac{\partial \ur}{\partial t}&=&-i\hhh\ur-\hhr\ur\\
&=&-\left(i\hhh+\hhr\right)\ur
\end{eqnarray}
We call \hhh \ the Hamiltonian superoperator and \hhr \ the relaxation superoperator. Exponentiating the generator gives us the superpropagator
\begin{equation}
\hhp(t)=e^{-\left(i\hhh+\hhr\right)t}\label{prop}.
\end{equation}
We can manipulate \eref{prop} to isolate the relaxation superoperator:
\begin{equation}
\hhr=-i\hhh-\frac{1}{t}\rm{ln}\left(\hhp\right) \label{riso}
\end{equation}

An alternative method of deriving \hhr \ from estimates of \hhp \ and \hhh \ is to estimate the derivative at $t=0$ of
\begin{equation}
e^{\frac{t}{2}i\hhh}e^{-\left(i\hhh+\hhr\right)t}e^{\frac{t}{2}i\hhh}.
\end{equation}
Here we need to invoke the symmetric Baker-Campbell-Hausdorff formula, for arbitrary operators $A$ and $B$,
\begin{equation}
e^{\frac{t}{2}A} e^{tB} e^{\frac{t}{2}A}=e^{\left(t(A+B)-\frac{1}{24}t^3\left[A+2B,\left[A,B\right]\right]+O(t^5)\right)}.
\end{equation}
Making the identification
\begin{eqnarray}
A&=&i\hhh, \\
B&=&-\left(i\hhh+\hhr\right),
\end{eqnarray}
we get
\begin{equation}
e^{\frac{t}{2}i\hhh}e^{-\left(i\hhh+\hhr\right)t}e^{\frac{t}{2}i\hhh}=e^{-t\hhr}+O(t^3).\label{etr}
\end{equation}

Clearly, differentiating \eref{etr} at $t=0$ gives us \hhr,  as required. In order to obtain this differential we used a numerical
differentiation technique - Richardson extrapolation \cite{richardson,havel}.

To use this technique we require estimates of the propagator, $\hhp_m=\hhp(t_m)$, at time points $t_m=2^mt_1$. This alone, however, is not a
robust method for estimating \hhr. It magnifies the noise present in the estimate of $\hhp_1$ \cite{havel}. More importantly, there are no
constraints on the physicality (i.e. complete positivity) of the generator. This estimate of \hhr \ which we will call \hhre \ is instead used
as the \emph{starting point} for a constrained fit to the experimental data.

In order to search for a physical process which was close to the measured data we used a parameterisation based on the Gorini Kossakowski
Sudarshan master equation:
\begin{equation}
\hh{\mathcal{L}}=-i\left[H,\rho\right]+\frac{1}{2}\sum_{\alpha,\beta=1}^{d^2-1} a_{\alpha
\beta}\left(\left[F_{\alpha}\rho,F_{\beta}^{\dag}\right]+\left[F_{\alpha},\rho F_{\beta}^{\dag}\right]\right).\label{eqgorini2}
\end{equation}
Gorini et al. have proven \cite{gorini} that $\hh{\mathcal{L}}$ is the generator of a Markovian semigroup on $d$-dimensional Hilbert space if
and only if it can be written in the form \eref{eqgorini2}, where the GKS matrix $a_{\alpha \beta}$ is a $(d^2-1) \times (d^2-1)$  positive
(semidefinite) matrix and $\left\{F_{\alpha}\right\}$ is a linear basis of traceless operators on $\rho$. We will exploit the requirement that
\gks \ be positive in a way that is completely analogous to a technique we used in standard QPT. During process reconstruction we enforced
positivity of $\xxt$, and, by extension, the complete positivity of \Er.
Similarly \textbf{we will enforce complete positivity of the decoherence process \hhr \ by constraining the GKS matrix $a_{\alpha \beta}$ to be
positive}.

For one qubit we have a parameterisation in terms of 9 real numbers $x(i)$:
\begin{equation}
-\hhr(x(i))=\frac{1}{2}\sum_{\alpha,\beta=1}^{d^2-1} a_{\alpha
\beta}(x(i))\left(\left[F_{\alpha}\rho,F_{\beta}^{\dag}\right]+\left[F_{\alpha},\rho F_{\beta}^{\dag}\right]\right)\label{linpars}
\end{equation}
where
\begin{equation}
\gks(x(i))=\mathbb{X}^{\dag}\mathbb{X} \qquad \mathbb{X}(x)=\left(%
\begin{array}{ccc}
  x(1) & 0 & 0  \\
  x(4)+ix(5) & x(2) & 0  \\
  x(8)+ix(9) & x(6)+ix(7) & x(3) \\
\end{array}%
\right).\label{linchol}
\end{equation}

There is considerable freedom in which basis $\left\{F_{\alpha}\right\}$ to use but we chose to use generators for SU(d), rescaled to be trace
orthonormal:
\begin{eqnarray}
F_1&=&\frac{1}{\sqrt{2}}\left(%
\begin{array}{cc}
  0 & 1 \\
  1 & 0 \\
\end{array}%
\right),\\
F_2&=&\frac{1}{\sqrt{2}}\left(%
\begin{array}{cc}
  0 & -i \\
  i & 0 \\
\end{array}%
\right),\\
F_3&=&\frac{1}{\sqrt{2}}\left(%
\begin{array}{cc}
  1 & 0 \\
  0 & -1 \\
\end{array}%
\right).
\end{eqnarray}

A numerical search, using Matlab's\textregistered \ fminsearch algorithm, was used to find the minimum of
\begin{equation}
\Delta(\hhr(x))=\sum_{m=1}^M \left|e^{-\left(i\hhh+\hhr(x)\right)t_m}-\hhp_m\right|^2 \label{rsearch}
\end{equation}
where $\hhp_m$ are the experimentally determined propagators, for evolution time $t_m$ i.e.
\begin{equation}
\hh{\mathcal{P}_m}=\left[\left(%
\begin{array}{c}
    \\
  \underline{\ket{\rho_1(t_m)}}\\
   \\
\end{array}%
\right)\left(%
\begin{array}{c}
    \\
  \underline{\ket{\rho_2(t_m)}}\\
   \\
\end{array}%
\right)\left(%
\begin{array}{c}
    \\
  \underline{\ket{\rho_3(t_m)}}\\
   \\
\end{array}%
\right)\left(%
\begin{array}{c}
    \\
  \underline{\ket{\rho_4(t_m)}}\\
   \\
\end{array}%
\right)\right]
\end{equation}
where
\begin{equation}
\underline{\ket{\rho_1(0)}}=\left(\begin{array}{c}
  1 \\
  0 \\
  0 \\
  0 \\
\end{array}\right), \ \
\underline{\ket{\rho_2(0)}}=\left(\begin{array}{c}
  0 \\
  1 \\
  0 \\
  0 \\
\end{array}\right), \ \
\underline{\ket{\rho_3(0)}}=\left(\begin{array}{c}
  0 \\
  0 \\
  1 \\
  0 \\
\end{array}\right), \ \
\underline{\ket{\rho_4(0)}}=\left(\begin{array}{c}
  0 \\
  0 \\
  0 \\
  1 \\
\end{array}\right).
\end{equation}

\begin{figure}[!ht]
    \begin{center}
    \subfigure{\epsfig{file=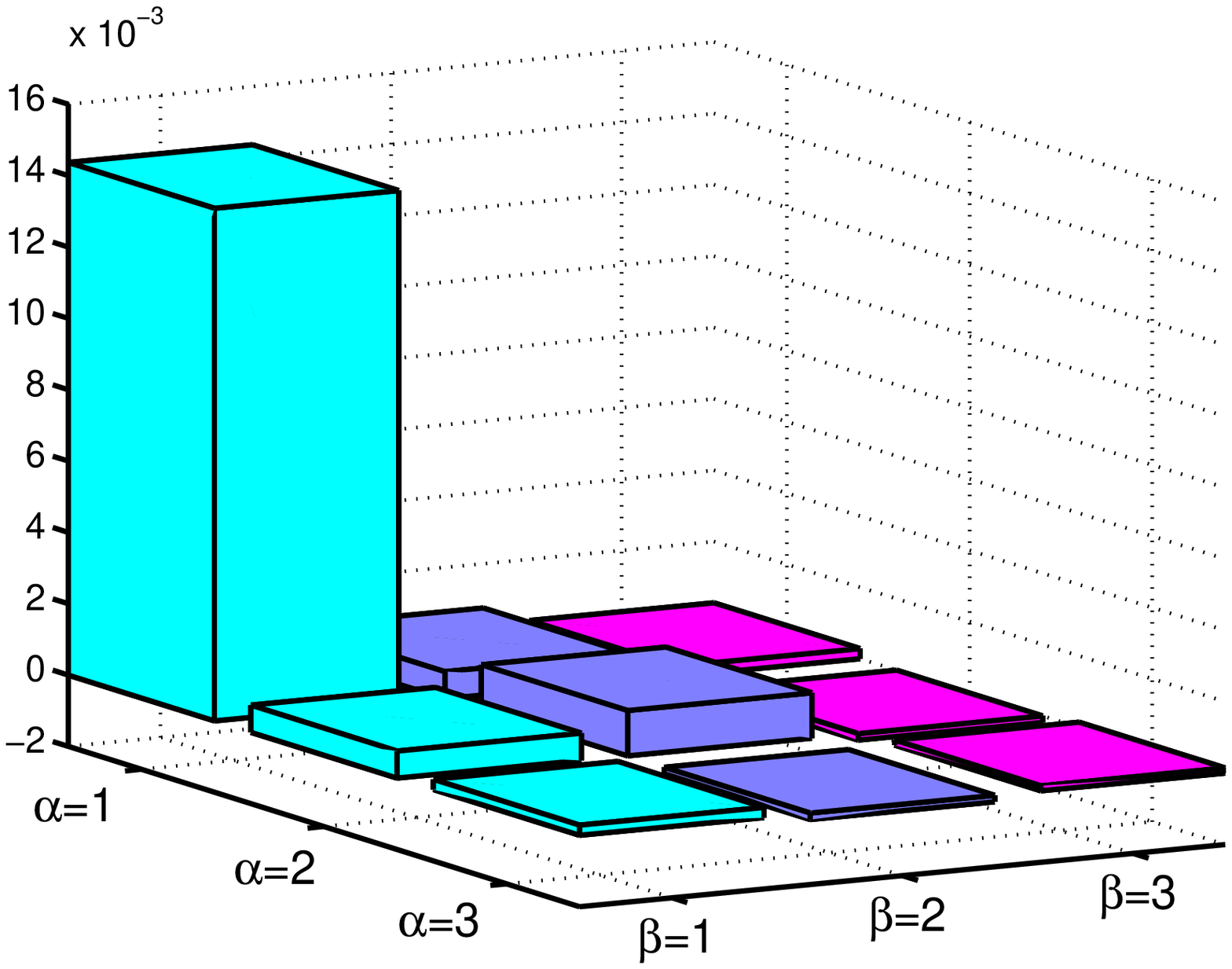, height=43mm}}{{(a)}}
    \subfigure{\epsfig{file=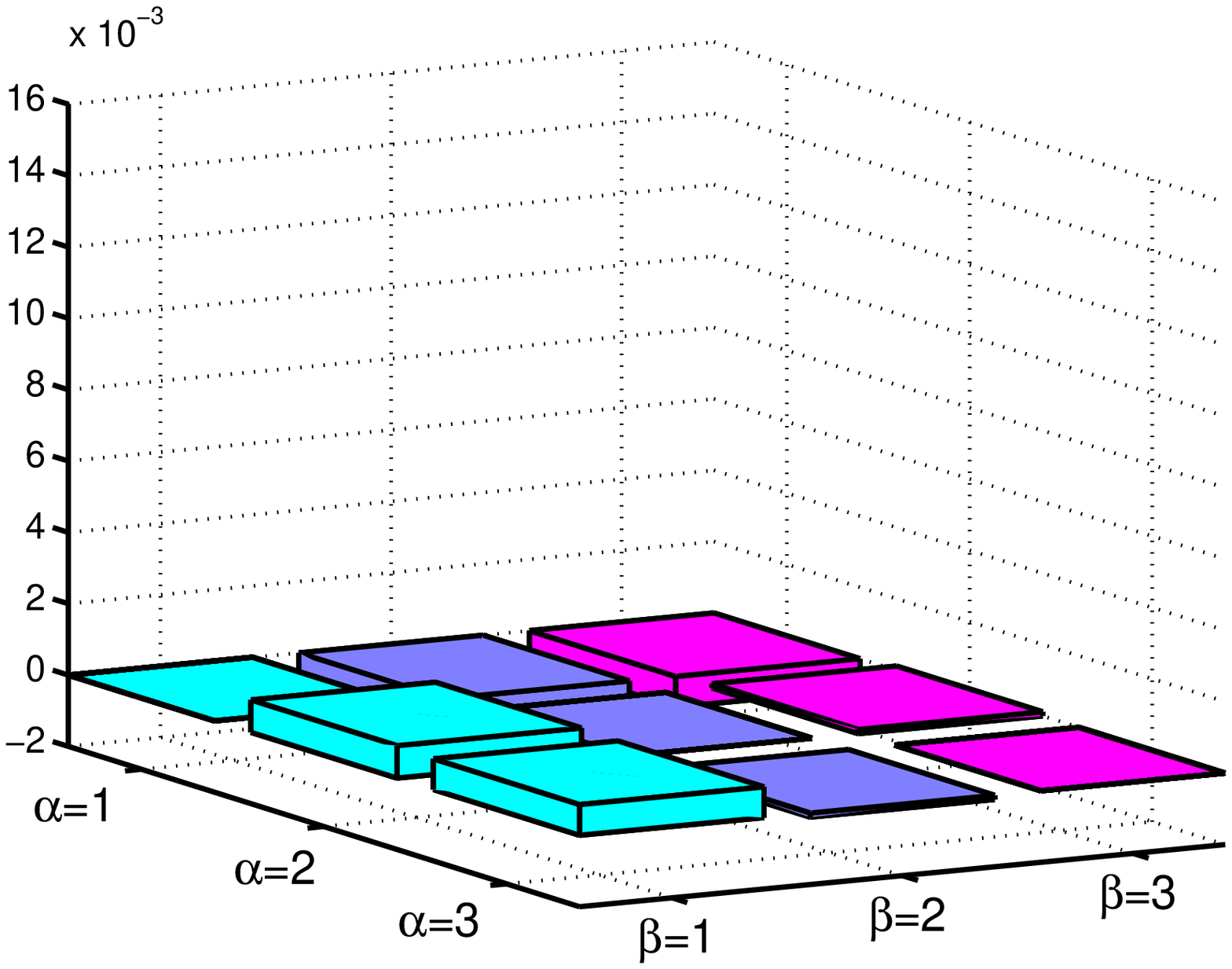, height=43mm}}{{(b)}}
\caption{\label{gksmatrix}GKS matrix \gks: the initial starting point for minimization of \eref{rsearch}  (a) Real components of \gks \ (b)
Imaginary components of \gks.}
\end{center}
\end{figure}

\begin{figure}[!ht]
    \begin{center}
    \subfigure{\epsfig{file=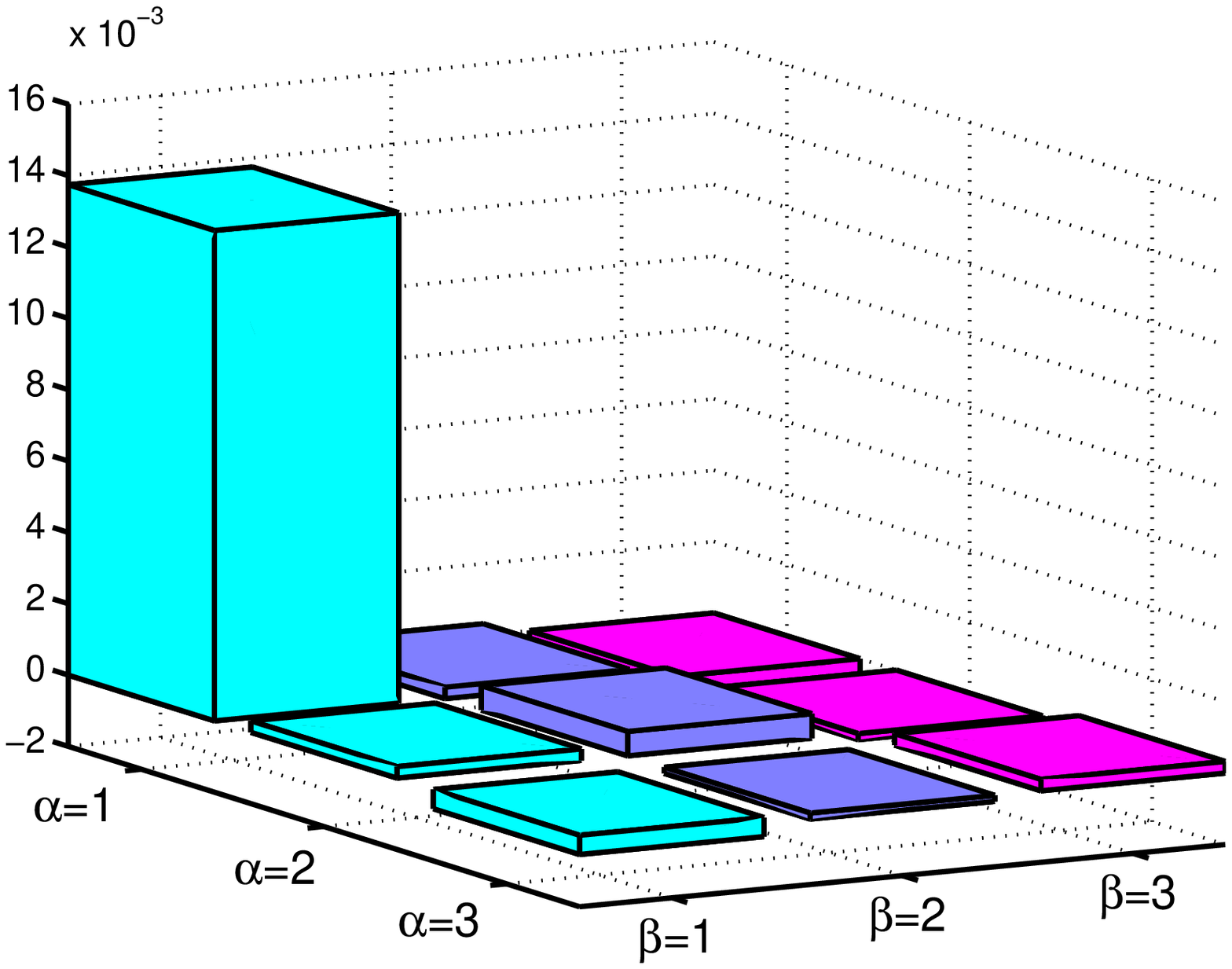, height=43mm}}{{(a)}}
    \subfigure{\epsfig{file=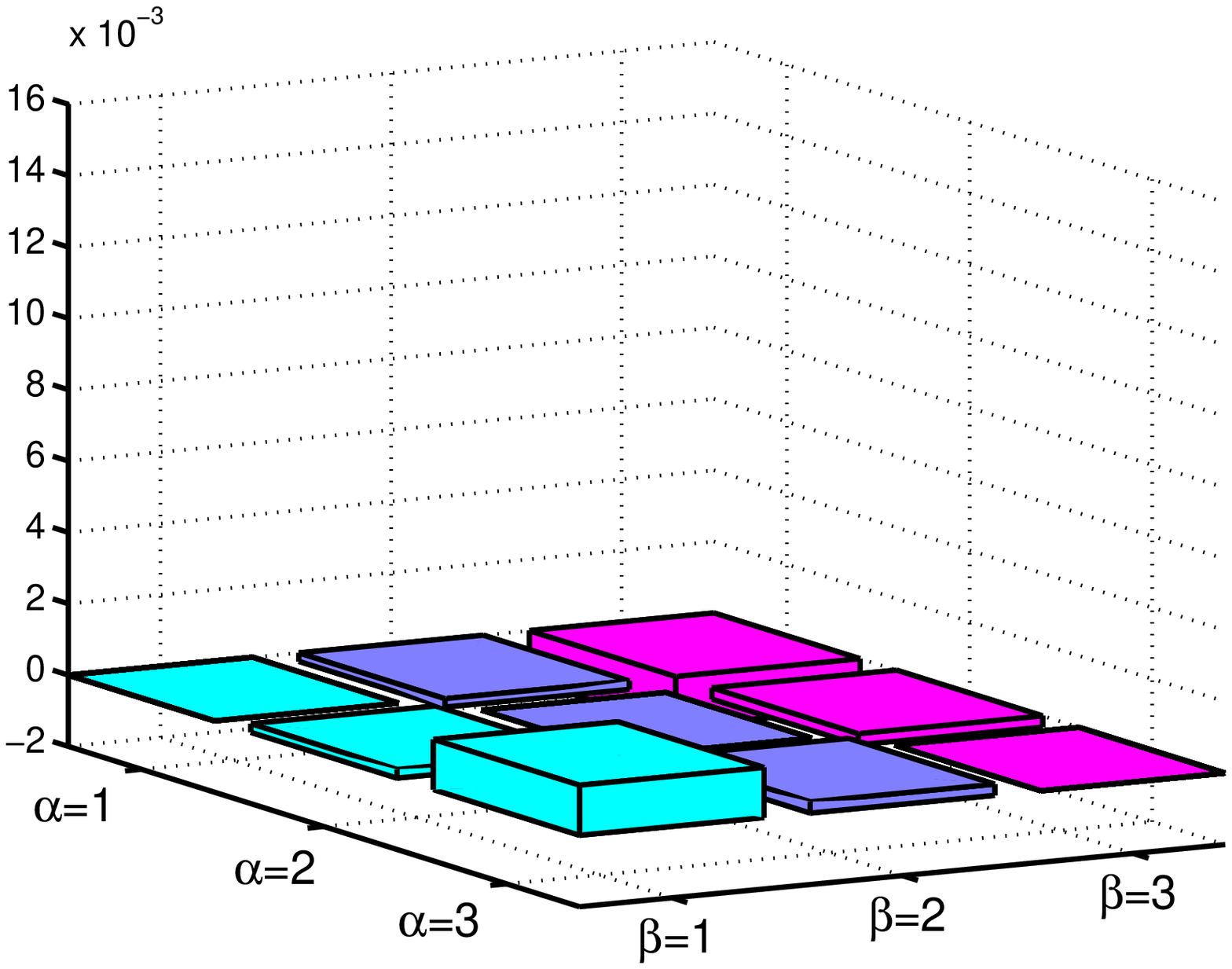, height=43mm}}{{(b)}}
\caption{\label{tgksmatrix}GKS matrix \tgks: the matrix which minimizes \eref{rsearch} (a) Real components of \tgks \ (b) Imaginary components
of \tgks.}
\end{center}
\end{figure}

\begin{figure}[!ht]
    \begin{center}
    \subfigure{\epsfig{file=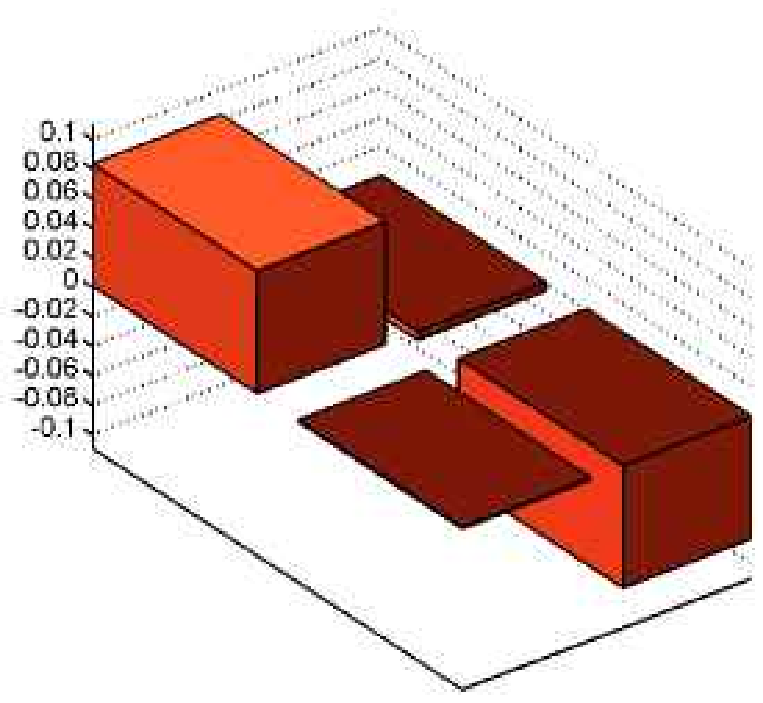, height=43mm}}{{(a)}}
    \subfigure{\epsfig{file=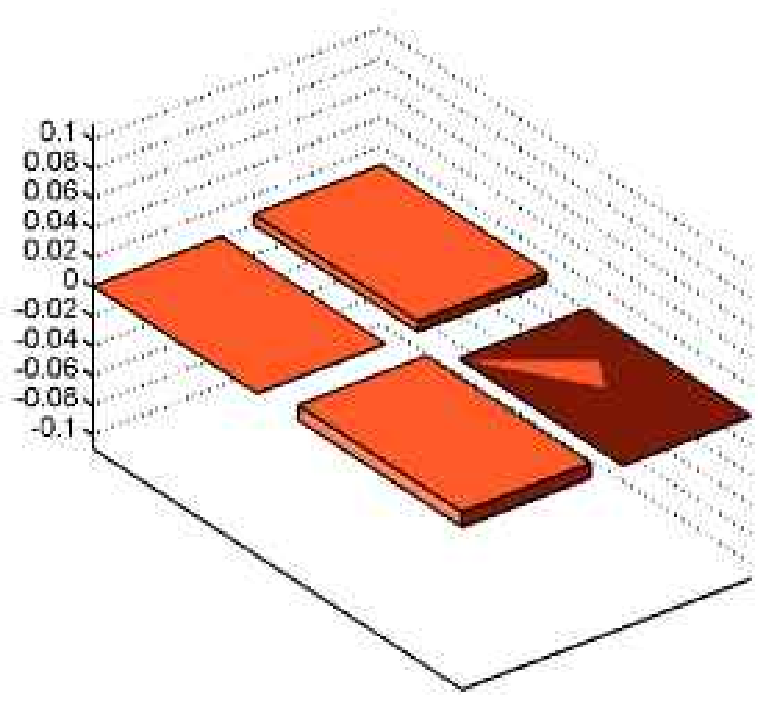, height=43mm}}{{(b)}}
\caption{\label{pL1}Lindblad operator $L_1$, Relative contribution $\approx 94.4\%$: (a) Real components of $L_1$. (b) Imaginary components of
$L_1$.}
\end{center}
\begin{center}\protect\begin{tabular}{|c|c|}
  \hline
   \\
    $L_1=\left(%
\begin{array}{cc}
   0.0829 - 0.0000i & -0.0056 - 0.0071i\\
  -0.0011 + 0.0101i & -0.0829 + 0.0000i\\
\end{array}%
\right)$ \\
\\
  \hline
\end{tabular}\end{center} \label{L1}
\end{figure}

\begin{figure}[!ht]
    \begin{center}
    \subfigure{\epsfig{file=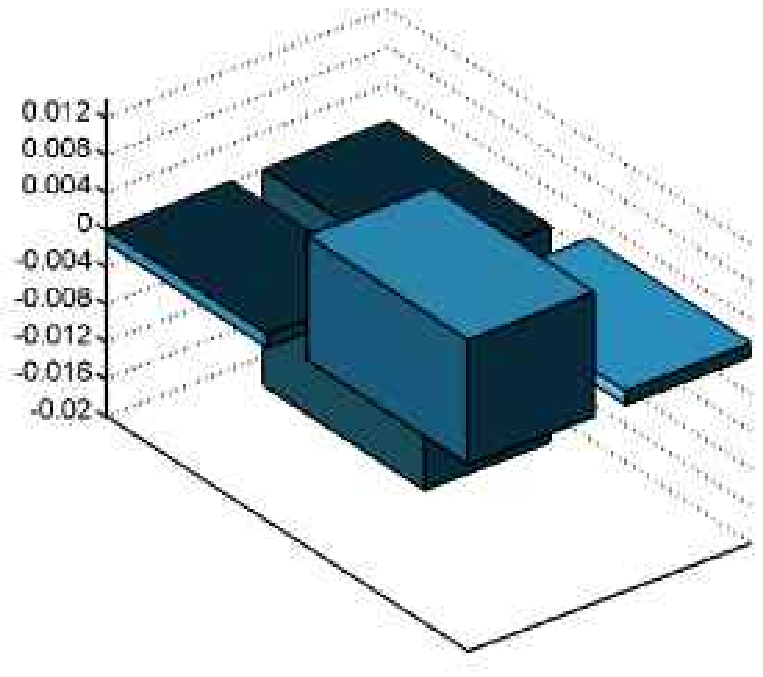, height=43mm}}{{(a)}}
    \subfigure{\epsfig{file=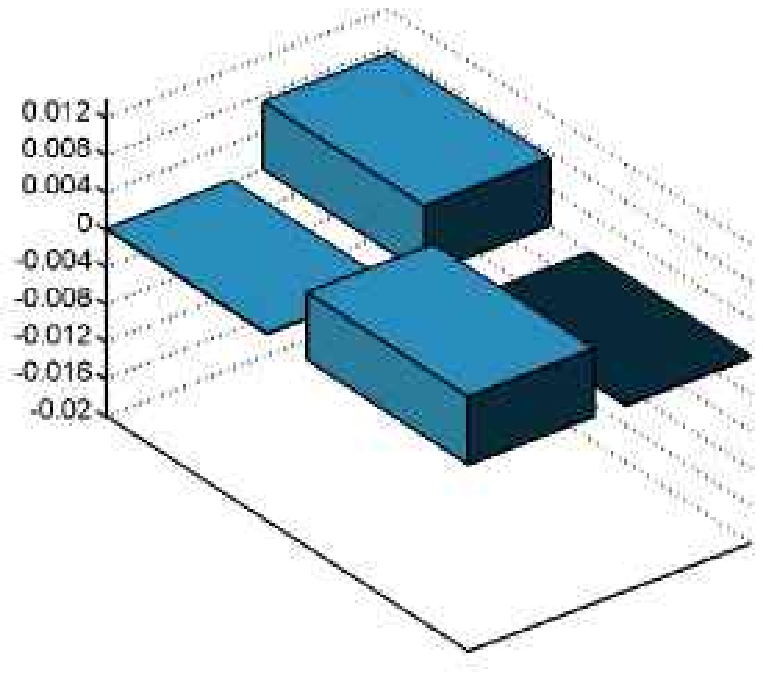, height=43mm}}{{(b)}}
\caption{\label{pL2}Lindblad operator $L_2$, Relative contribution $\approx 5.6\%$: (a) Real components of $L_2$. (b) Imaginary components of
$L_2$.}
\end{center}
\begin{center}\protect\begin{tabular}{|c|c|}
  \hline
   \\
    $L_2=\left(%
\begin{array}{cc}
  -0.0014 + 0.0001i & -0.0232 + 0.0072i\\
   0.0134 + 0.0072i &  0.0014 - 0.0001i\\
\end{array}%
\right)$ \\
\\
  \hline
\end{tabular}\end{center} \label{L2}
\end{figure}

\begin{figure}[!ht]
    \begin{center}
    \subfigure{\epsfig{file=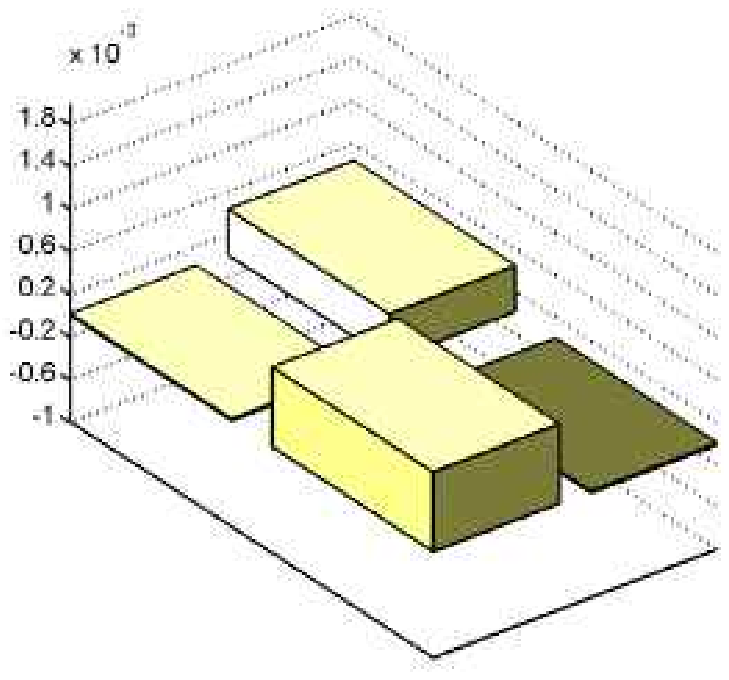, height=43mm}}{{(a)}}
    \subfigure{\epsfig{file=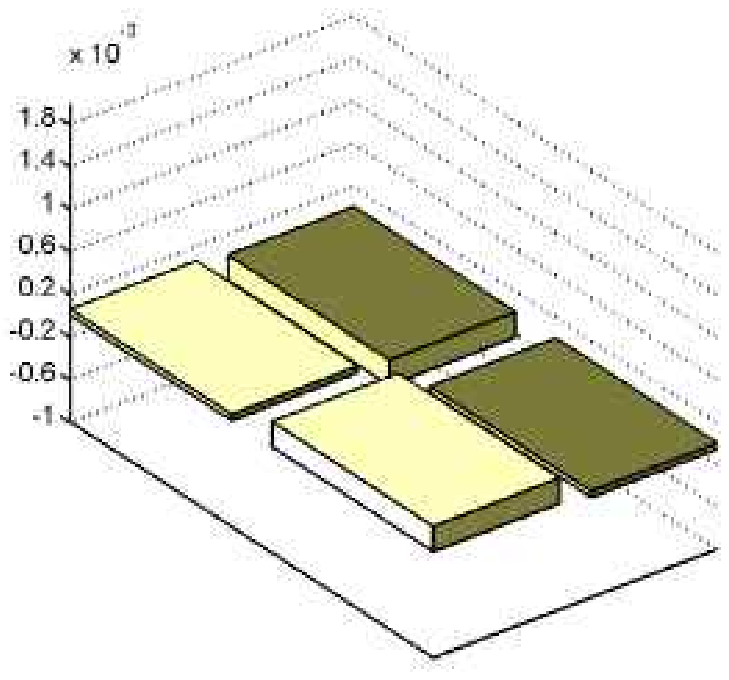, height=43mm}}{{(b)}}
\caption{\label{pL3}Lindblad operator $L_3$, Relative contribution $\approx \left(1\times10^{-4}\right)\%$: (a) Real components of $L_3$. (b)
Imaginary components of $L_3$.}
\end{center}
\begin{center}\protect\begin{tabular}{|c|c|}
  \hline
   \\
    $L_3=1\times10^{-3}\left(%
\begin{array}{cc}
  0.0151 + 0.0577i &  0.4324 - 0.2377i \\
   0.7500 + 0.2377i &  -0.0151 - 0.0577i \\
\end{array}%
\right)$ \\
\\
  \hline
\end{tabular}\end{center} \label{L3}
\end{figure}

\begin{figure}[h]
    \begin{center}
    \epsfig{file=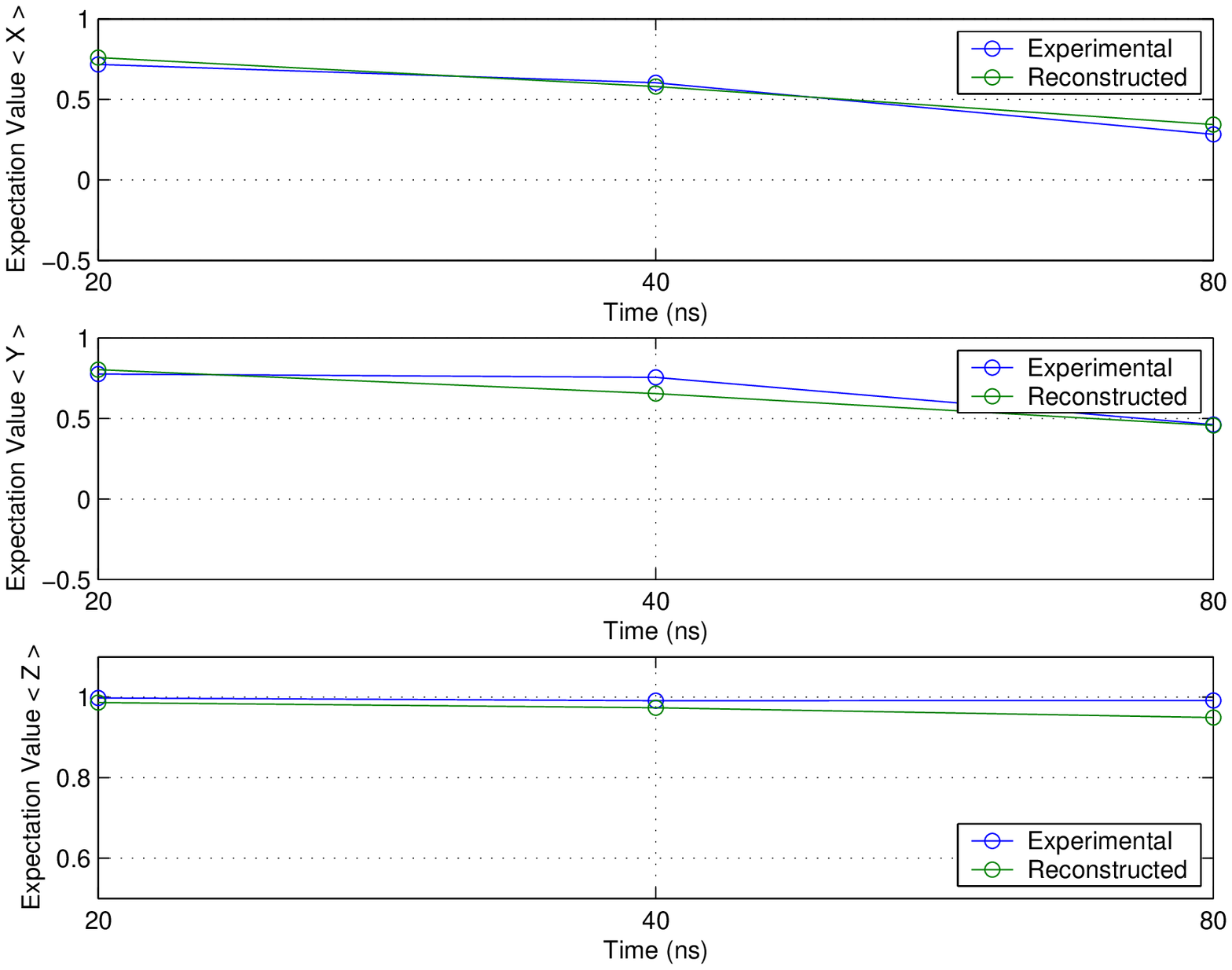,height=110mm, width=120mm}
    \caption{\label{lindbladtriple} Comparison of experimental expectation values at timepoints $t_m$ with expectation values predicted by Markovian master equation. The time is plotted on a logarithmic scale for clarity.}
    \end{center}
\end{figure}

We label the GKS matrix which most closely fits the data, while remaining physical, as $\tgks$ (see figures \ref{gksmatrix} and
\ref{tgksmatrix}). To derive the Lindblad operators $\{L _k\}$ from $\tgks$ we first diagonalise $\tgks$ with a unitary $U^{\dag}$,
\begin{equation}
\widetilde{a_{mn}}=\sum_{xy} U_{mx}d_{x} \delta_{xy} U^{*}_{ny},
\end{equation}
where $d_i$ are the eigenvalues of $\tgks$. The Lindblad operators can then be obtained by
\begin{equation}
L_i=\sqrt{d_i}\sum_j U_{ji}F_j.
\end{equation}
These are depicted in figures \ref{pL1} to \ref{pL3}. The expectation values $<\hat{\sigma_x}>$,$<\hat{\sigma_y}>$ and $<\hat{\sigma_z}>$ for
both the experimental and the reconstructed (Markovian, completely positive) process are compared in figure \ref{lindbladtriple}. It is natural
now to ask given the estimations for $L_i$, whether one can deduce any information regarding the physical decoherence processes responsible.
Unfortunately, this is not generally possible. The most general Markovian master equation for a qubit can be fully described by three Linblad
operators. However this compressed description often will not correspond to the most appropriate physical description of the decoherence. It can
be argued however, that in most situations, e.g. decoherence free subspace investigation, decoupling pulse generation etc., all the relevant
information is contained in the compressed Lindblad description.

The relative contribution of each Lindblad operator, $L_i$, to the overall decoherence process can be quantified using their Frobenius norms:
\begin{equation}
\rm{Relative \ Contribution}(L_i)=\frac{|L_i|_{Fro}^2}{\sum_{j=1}^{d^2-1}|L_j|^2_{Fro}}.
\end{equation}

In summary the technique we used for ascertaining the relaxation superoperator \hhr \ is as follows:
\begin{description}
\item $\bullet$ Initialisation via optical pumping and appropriate ESR pulses to create a complete basis of input states $\rho_i$.

\item $\bullet$ Each $\rho_i$ was left to decohere for a sequence of times $t_m=2^mt_1$ i.e. \{20ns, 40ns, 80ns\}.

\item $\bullet$ State tomography was used to determine the output state $\rho_i(t_m)$ after each decoherence time $t_m$. The results were used
to construct the matrix superpropagators $\hhp_m$.

\item $\bullet$ Using Richardson extrapolation, an estimate of the generator, \hhre, \ was extracted from the measured $\hhp_m$ and used as a starting point for the fitting procedure \eref{rsearch}.

\item $\bullet$ The minimisation of \eref{rsearch} produced a Markovian generator, $\hat{\hat{\widetilde{\mathcal{R}}}}$, which was physically valid and best fit the measured data at
a sequence of timepoints $t_m$.

\item $\bullet$ The GKS matrix which minimised \eref{rsearch}, \tgks, was then diagonalised in order to find the Lindblad
operators $\left\{L_k\right\}$.

\end{description}

\subsection{Markovian process tomography results}

\ \vspace{30mm} \ \ \ \ \ \ \ \ \ \ \ \ \

\section{Summary}
We have presented the first quantum process tomographic analysis of  an individual single solid-state qubit, a Nitrogen Vacancy centre in
Diamond.  This analysis is only possible due to the enormous advances made in recent years in single-molecule spectroscopy
\cite{thir,fift,sixt}, where the resultant ODMR technique provides us here with high-fidelity single-qubit readout. As experimental refinements
and technological advances improve the readout and control of this system, QPT will become an even more important diagnostic tool. For example,
improvement of the $m_s=0$ polarisation by optical pumping and increased accuracy of rotations by the use of composite pulse sequences, are both
feasible short-term goals. If states can be prepared, controlled and read out with very high accuracy, then any deviations from Markovian
dynamics (e.g. the disparity in expectation values, between experimental and reconstructed processes apparent in figure \ref{lindbladtriple})
cannot be dismissed as noise and the nature of the non-Markovian environment should be investigated. If two qubit gates can be performed in this
system then QPT can be used to verify the robustness of encoded information in decoherence free subspaces. As quantum devices develop and
increase in size, the task of ``debugging" the device, or actively identifying the noise present in the device, will pose significant
challenges. The work presented here represents an initial step towards the testing of quantum devices in solid-state.

\end{document}